\documentclass[12pt]{article}

\usepackage{times}

\usepackage{color}
\usepackage{xcolor}
\usepackage{graphicx}
\usepackage{amsmath,amssymb}
\usepackage{amsthm}
\usepackage{xr}
\usepackage{float}
\usepackage{longtable}

\usepackage{xr}
\usepackage{longtable}
 
\newenvironment{sciabstract}{%
\begin{quote} \bf}
{\end{quote}}

\title{Automation and occupational mobility:\\
A data-driven network model\\
\vspace{4mm}
\vspace{2mm}
}

\author
{R. Maria del Rio-Chanona ,$^{1, 2\ast}$ Penny Mealy,$^{1,4,5}$  Mariano Beguerisse-D\'iaz,$^{2}$\\  Fran\c{c}ois Lafond,$^{1, 3, 4}$ and 
 J. Doyne Farmer$^{1,2,6}$\\
\\
\normalsize{$^{1}$Institute for New Economic Thinking at the Oxford Martin School, University of Oxford}\\
\normalsize{$^{2}$Mathematical Institute, University of Oxford}\\
\normalsize{$^{3}$Oxford Martin School Programme on Technological and Economic Change}\\
\normalsize{$^{4}$School of Geography and Environment, University of Oxford}\\
\normalsize{$^5$ Bennett Institute for Public Policy, University of Cambridge }
\normalsize{$^{6}$Santa Fe Institute,}\\
\vspace{4mm}
\small{$^\ast$To whom correspondence should be addressed (rita.delriochanona@maths.ox.ac.uk).}
}

\date{}

\begin{document} 


\baselineskip16pt
\maketitle 
    
\begin{sciabstract}
The potential impact of automation on the labor market is a topic that has generated significant interest and concern amongst scholars, policymakers, and the broader public. A number of studies have estimated occupation-specific risk profiles by examining the automatability of associated skills and tasks. However, relatively little work has sought to take a more holistic view on the process of labor reallocation and how employment prospects are impacted as displaced workers transition into new jobs. In this paper, we develop a new data-driven model to analyze how workers move through an empirically derived occupational mobility network in response to automation scenarios which increase labor demand for some occupations and decrease it for others. At the macro level, our model reproduces a key stylized fact in the labor market known as the Beveridge curve and provides new insights for explaining the curve’s counter-clockwise cyclicality. At the micro level, our model provides occupation-specific estimates of changes in short and long-term unemployment corresponding to a given automation shock. We find that the network structure plays an important role in determining unemployment levels, with occupations in particular areas of the network having very few job transition opportunities. Such insights could be fruitfully applied to help design more efficient and effective policies aimed at helping workers adapt to the changing nature of the labor market.
\end{sciabstract}

\section*{Introduction}
In response to widespread concern about the potential impact of automation on the labor market \cite{national2017information,brynjolfsson2017artificial,autor2003skill,autor2006polarization,david2015there}, significant effort has been devoted towards analyzing how susceptible a given occupation is to computerisation \cite{frey2017future, brynjolfsson2018SML,arntz2016risk}. However, studies that estimate the likelihood of a robot `stealing' a particular job only provide part of the picture. Consider, for example, the job security faced by a statistical technician vs. a childcare worker. Estimates developed by Frey and Osborne \cite{frey2017future} suggest that statistical technicians are more likely than childcare workers to be replaced by software technology. However, should such forecasts eventuate, and statistical technicians find themselves out of a job,  their existing skills could allow them to transition into a range of `safer’ occupations with lower automation risk and growing demand. In contrast, while childcare workers may not experience a $direct$ threat from computerization, their employment prospects may nonetheless still be jeopardized. As automation displaces people in other occupations, many of these workers could have the requisite skills to become childcare workers and may consequently provide an $indirect$ threat to the job security of existing childcare workers. Thus, even though the immediate risk of automation is predicted to be larger for statistical technicians, accounting for possible occupational transitions and labor demand reallocation could see childcare workers facing a greater risk of unemployment. 

Building on the rich body of literature that has demonstrated the importance of modeling labor flows using networks and agent-based models \cite{schmutte2014free,nimczik2017job,guerrero2013employment,lopez2015network,axtell2019frictional,dworkin2019network,neffke2017inter,diodato2014resilience,alabdulkareem2018unpacking,michael2018agent,goudet2017worksim}, this paper develops a new data-driven labor-market model to study these important but overlooked indirect labor displacement effects. Central to our model is an empirically derived \emph{occupational mobility network}, in which nodes are different occupations and edges correspond to the probability that workers transition between them. The overall structure of this network influences the efficiency with which workers are reallocated across occupations following a shift in relative labor demand.

To explore the potential impacts of automation on the labor market, we impose an automation ‘shock’ that, over the years, decreases demand for labor in some occupations an increases demand in others. Using an agent-based model, we study the associated aggregate and occupation-specific unemployment dynamics as a function of time. While we analyze the results for only two automation shock scenarios based on estimates developed by Frey and Osborne \cite{frey2017future} and Brynjolfsson et al. \cite{brynjolfsson2018SML}, our model is quite general and can be used to study a range of different labor market shocks. 

We model the resulting process of labor reallocation as a stochastic process with discrete time steps. During each time step, occupations open vacancies and separate (fire) workers, unemployed workers apply for a new job, and vacancies and job applications are matched. We model an out-of-equilibrium economy and focus on the transient dynamics during which the labor market re-adjusts to a new steady-state equilibrium. In addition to simulations, we also derive a representation of the model as a deterministic dynamical system that, in the limit of a large number of workers, predicts the expected behavior of the model simulations. Using the dynamical system equations allows us to obtain deeper insights into the mechanics of the model, dramatically speed up analyzes of the model’s dynamics with little loss of accuracy, and obtain a practically complete analytical characterization of the model in particular cases.

Imposing Frey and Osborne’s automation scenario in our model generates a substantial increase in aggregate unemployment for a period of around a decade. We analyze the impacts on both short-term and long-term unemployment (> 27 weeks). Not surprisingly, occupations that are at higher risk of automation tend to be affected most. However, the occupational mobility network plays an important role in determining aggregate and occupation-specific unemployment levels. 

Specifically, we show that restrictions on worker movements imposed by the occupational mobility network generate significant labor market mismatch. In some areas of the network, many workers can be competing for very few vacancies. At the same time, occupations in other areas can have job vacancies that are left unfilled for long periods. In comparison to a labor market with no mobility restrictions, the occupational mobility network structure increases unemployment by roughly $25\%$. We also show that occupations with the same level of \emph{ex ante} automation risk can end up with markedly different unemployment levels. For example, while dispatchers and pharmacy aids are both estimated to have a computerization probability of 0.72, dispatchers face a $19\%$ increase in long-term unemployment, while long term unemployment for pharmacy aids \emph{decreases} by about the same amount. 

Our model also provides unique insights into the Beveridge Curve, which is a well-known negative empirical relationship between the unemployment rate and the vacancy rate \cite{beveridge2014full}. Typically, when vacancies open up, unemployment goes down. We show that under calibrated parameter values, our model is able to reproduce the empirical Beveridge Curve during the most recent US business cycle, and supports the hypothesis that business cycles alone can cause the counter-clockwise cycling behavior of the curve \cite{elsby2015beveridge,diamond2015shifts,sniekers2018persistence}.

Our results have important implications for the design of policies aimed at helping workers best prepare and adapt to the changing nature of the labor market. More nuanced insights into employment impacts associated with automation could help improve the effectiveness of worker retraining schemes. For example, rather than only considering workers’ current occupation’s susceptibility to automation, skill development programs could be more efficiently targeted towards workers in occupations that are likely to face longer spells of unemployment \cite{dworkin2019network}. Further, a better understanding of the mechanisms underpinning the Beveridge curve could help policymakers mitigate adverse employment impacts of business cycles and accelerate the recovery process.

\section*{Results} \label{sec:results}

\subsection*{The occupational mobility network}
\begin{figure}
\begin{center}
\includegraphics[width=0.5\linewidth]{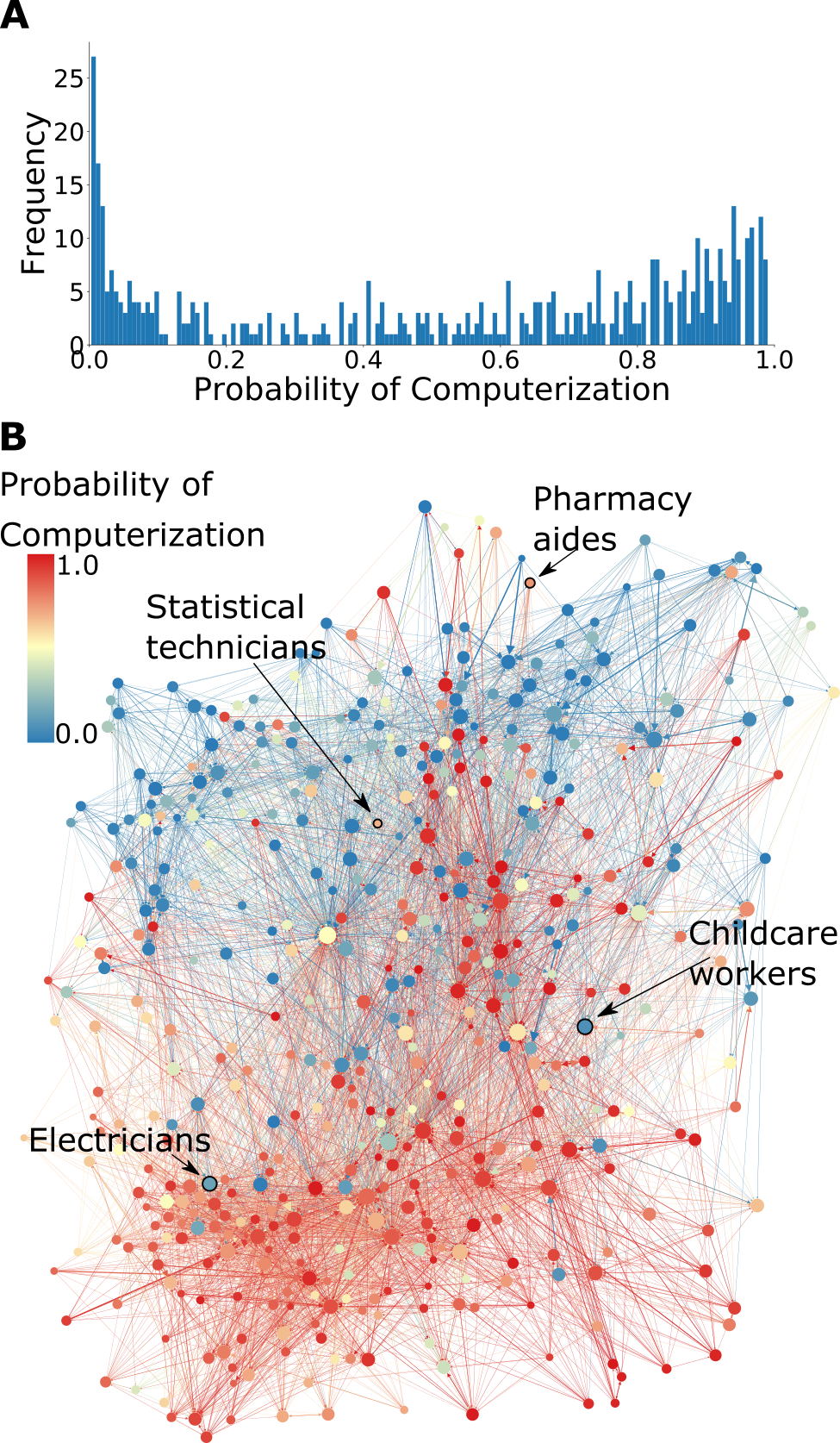}%
\caption{\textbf{Estimates of automatability in the occupational mobility network.} Panel (\textbf{A}) is a histogram of the probability of computerization for different occupations as estimated by Frey and Osborne \cite{frey2017future}. Noticeably, the probability of computerization has a bimodal distribution. Panel (\textbf{B}) shows the occupational mobility network, where nodes represent occupations and links represent possible worker transitions between occupations. The color of the nodes indicates the estimated probability of computerization. Red nodes have high automatability and blue nodes have low automatability. The size of the nodes indicates the number of employees in each occupation.}
\label{fig:network_fo}
\end{center}
\end{figure}

We first construct an occupational mobility network representing the ease with which a worker can transition between occupations.  Here we follow the work of  Mealy et al. \cite{mealy2018} and construct the network based on data on occupational transitions in the United States between 2010--2017.  In this network nodes are occupations, and the weights of the edges are proportional to the probability that a worker transitions between occupations. The resulting network is weighted and directed with $n=464$ nodes (see Fig. \ref{fig:network_fo}). The network also has self-loops, since workers often remain in the same occupation when they change jobs. We represent the network by its adjacency matrix $A$, with elements
\begin{equation}
 A_{ij}= 
  \begin{cases} 
   r & \text{if } i = j  \\
   (1 - r)P_{ij}  & \text{if } i \neq j,
  \end{cases}
\end{equation}
where the indices $i$ and $j$ label the $n$ possible occupations.  $r$ is the weight of the self-loops: It is the probability that a worker who is changing jobs applies to a job vacancy in her original occupation. $P_{ij}$ is the empirical probability that a worker transitioning out of occupation $i$ moves to occupation $j$. For details on how we compute $P_{ij}$ we refer the reader to the \textit{Methods} section.  In this paper we assume that $A_{ij}$ is fixed in time -- edges do not change, and no nodes are removed or added. 

As shown in Fig \ref{fig:network_fo}, the set of possible job transitions has a rich network structure \cite{mealy2018}.  This reinforces recent studies that have shown that occupational mobility is significantly more restricted than is commonly assumed in most labor-market models \cite{autor2013task, petrongolo2001looking}.  Fig. \ref{fig:network_fo} also shows how estimates of the automatability of occupations are distributed across the occupational mobility network assuming estimates by Frey and Osborne. 

\subsection*{A network model of the labor market}
\begin{figure}
\begin{center}
\includegraphics[width=0.5\linewidth]{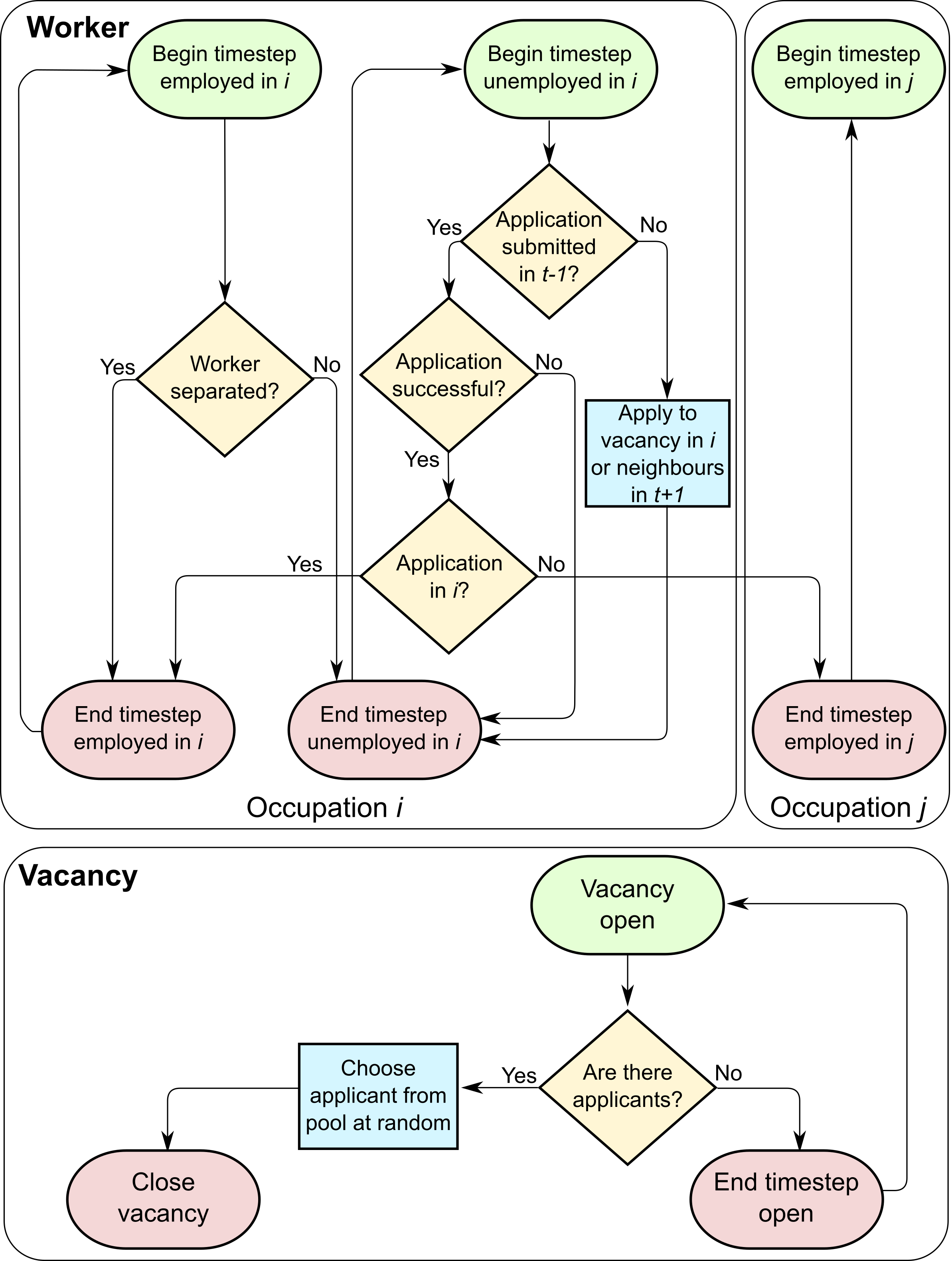}
\caption{\textbf{Flow chart illustrating the possible transitions of workers and job vacancies during a given time step.} {\it Top}: transitions of a worker. {\it Bottom}: The transitions of a job vacancy.  Note that vacancies created in the current time step do not accept job applications until the following time step.}
\label{fig:diagram}
\end{center}
\end{figure}

Our model is designed to understand the dynamics of unemployment at the occupation level. The flow of workers on the network is described by a set of discrete-time stochastic processes for employment, unemployment, and vacancies in each occupation $i$.  In this model, we assume that workers are perfectly geographically mobile, and we neglect wage pressure.  The set of possible occupations is fixed, and the occupation of a worker is defined as the occupation in which she was last employed. At any given time $t$ the number of workers employed in occupation $i$ is $e_{i,t}$, the number of unemployed workers is $u_{i,t}$, and the number of job vacancies  is $v_{i, t}$. The number of workers that are separated (i.e. fired) is $\omega_{i,t}$ and the number of vacancies is $\nu_{i,t}$.   The labor flow $f_{ij, t+1}$ is the number of workers hired in occupation $j$ who were previously unemployed in occupation $i$.   
The resulting set of stochastic processes can be described by equations, 
\begin{align}
 && e_{i,t + 1}  &=  e_{i,t}  - \ \underbrace{  \omega_{i, t + 1} }_\textrm{separated workers} \ +  \underbrace{\sum_j f_{ji, t +1}}_\textrm{hired workers}
  \label{eq:e_t_exact}&\\
 &&  u_{i,t + 1}  &=  u_{i, t} +  \underbrace{ \omega_{i, t+1} }_\textrm{separated workers} \ - \underbrace{\sum_j f_{ij, t+1}}_\textrm{transitioning workers} 
 \label{eq:u_t_exact}&\\
 &&  v_{i,t + 1}  &= v_{i, t} +   \underbrace{ \nu_{i, t+1} }_\textrm{opened vacancies} \ -  \underbrace{\sum_j f_{ji, t+1}}_\textrm{hired workers}.
\label{eq:v_t_exact} 
\end{align} 
These equations express conservation laws stating that the change in each variable is equal to the difference between inflow and outflow. Eq. (\ref{eq:e_t_exact}) states that the change in employment is equal to the number of workers that are hired minus the number of workers that are separated. Similarly, Eq. (\ref{eq:u_t_exact}) states that the change in unemployment is equal to the number of workers who are separated minus the number who are hired. Finally, Eq. (\ref{eq:v_t_exact}) states that the number of vacancies is equal to the number of vacancies created minus the number of workers who are hired to fill them.  Fig. \ref{fig:diagram} is a flow chart that makes the transitions explicit from the perspective of a worker and from the perspective of a job vacancy. 

We denote occupation-specific variables by lower-case letters and aggregate quantities by upper-case letters (e.g., total unemployment is $U_t = \sum_i u_{i,t}$) and use bold font for vectors (i.e., the $i$th element of $\mathbf{u}_t$ is $u_{i, t}$). The time steps are chosen so that their duration is long enough for workers to transition between occupations, but too short for workers to change their employment status more than once. That is, a worker is not allowed to switch her status from employed to unemployed and then back to employed in a single time step. Likewise, a vacancy cannot be opened and filled within the same time step.

We assume that the number of separated workers $\omega_{i, t + 1}$ and the number of job openings $\nu_{i,  t+ 1}$ follow binomial processes of the form,
\begin{align}
  \label{eq:omega_dist}
    \omega_{i, t + 1} \sim \text{Bin}(e_{i,t}, \pi_{u,i,t}),\\
    \nu_{i, t + 1} \sim \text{Bin}(e_{i,t}, \pi_{v,i,t}),
    \label{eq:nu_dist}
\end{align}
where $\text{Bin}(m, p)$ denotes a binomial distribution with $m$ trials and success probability $p$.  The success probabilities $\pi_{u,i,t}$ and $\pi_{v,i,t}$ depend on the imbalance of supply and demand for labor and play a key role in the dynamics.  Because this is a bit complicated, we will first complete our overview of the model and return in a moment to specify $\pi_{u,i,t}$ and $\pi_{v,i,t}$.

The labor flow $f_{ij, t+1}$ depends on the structure of the occupational mobility network, the number of vacancies and unemployed workers, and the processes of job search and job matching. The search and matching process can be thought of as an urn problem. Imagine each worker has one ball with her name on it and each job vacancy is an urn\footnote{We assume, as other have done \cite{axtell2019frictional}, that workers send only one job application per time step. This facilitates the mathematical derivations.}. With probability $A_{ij}$ each worker in occupation $i$ picks an urn corresponding to occupation $j$ and places her ball in it. After all workers have placed their balls, a ball is drawn from each urn with uniform probability and the corresponding worker is hired. If a vacancy does not receive job applicants, it remains open on the next time step. 

The labor flow $f_{ij, t+1}$ is thus a stochastic variable that can be computed based on the fact that the probability that a worker makes a transition from occupation $i$ to occupation $j$ is equal to the probability $q_{ij,t+1}$ that she applies to $j$, times the probability $p_{ij,t+1}$ that her application is accepted.  The probability $q_{ij, t+1}$ that an unemployed worker in occupation $i$ applies to a vacancy in occupation $j$ is
\begin{equation}
 q_{ij, t+1} = \frac{v_{j, t} A_{ij}}{\sum_l v_{l,t} A_{il}}.
 \label{eq:q}
\end{equation}
This means the expected number of applications submitted from occupation $i$ to occupation $j$ is 
\begin{equation}
E[s_{ij,t+1}| u_{i,t}] = u_{i, t} q_{ij, t+1}.
 \label{eq:s}
\end{equation}
Since each unemployed worker sends one job application, for fixed $i$ the random variables $s_{ij,t+1}$ follow a multinomial distribution with $u_{i,t}$ trials and probabilities $q_{ij, t+1}$ for $j = 1,...,n$.    All vacancies that have applications hire one worker. However, some vacancies may lack applications, in which case no one is hired, and the job vacancy remains open.  As we will make explicit later, the probability $p_{ij,t+1}$ that an application is successful follows from the urn model \cite{stevens2007new}. 

\paragraph{Supply and demand for labor}
Workers move across the occupational mobility network in response to shifts in labor supply and demand.  This is determined by the success probability $\pi_{u,i,t}$ of the binomial process for separating workers in Eq.~(\ref{eq:omega_dist}) and the success probability $\pi_{v,i,t}$ for the binomial process for creating vacancies in Eq.~(\ref{eq:nu_dist}). We break each of these into two separate random processes.  The first is a {\it spontaneous process} (or state-independent) and the second is a {\it state-dependent process}.   

In the spontaneous process, workers are separated and vacancies are opened at random, independent of the state of the system.  For simplicity here we assume that the separation and opening rates are the same for all occupations. For any given occupation, the spontaneous probability that a given worker is separated at any given time is $\delta_u$, and the spontaneous probability that a vacancy opens is  $\delta_v$ times the number of workers in that occupation.

The state-dependent process accounts for imbalances in supply and demand.  This is done by adding an additional occupation-specific probability $\alpha_{u,i,t}$ that a worker from occupation $i$ is separated at time $t$ and an additional occupation-specific probability  $\alpha_{v,i,t}$ that a vacancy in occupation $i$ opens.  Both of these probabilities are functions of time, constructed to equilibrate supply and demand. 
The \emph{target} labor demand  $d_{i, t}^\dagger$ is the desired quantity of labor for occupation $i$ at time $t$.  This is imposed externally and allows us to impose automation shocks as a function of time.  The \emph{realized} labor demand, in contrast, is a time-dependent variable corresponding to the sum of the number of employed workers plus the number of job vacancies in a given occupation, i.e.
$$
   d_{i,t} = e_{i,t} + v_{i,t}.
$$
$\alpha_{u,i,t}$ and $\alpha_{v,i,t}$ satisfy the following conditions:
\begin{enumerate}
    \item If the realized labor demand of an occupation equals the target labor demand, i.e., $d_{i,t} - d_{i,t}^\dagger = 0$, then no adjustments are made, i.e. $\alpha_{u,i,t} = \alpha_{v,i,t} = 0$.
    \item $\alpha_{u,i,t}$ is an increasing function of $d_{i,t} - d_{i,t}^\dagger$. This condition guarantees that when the realized demand of an occupation is greater than the target demand, workers of that occupation are more likely to be separated and thus decrease the occupation's realized demand. Likewise, $\alpha_{v,i,t}$ is an increasing function of $d_{i,t}^\dagger - d_{i,t}$, so that when the realized demand is less than the target demand, more vacancies are likely to open in that occupations and thus increase the occupation's realized demand.
    \item $\alpha_{u,i,t}$ and $\alpha_{v,i,t}$ are probabilities and lie in the interval $[0,1]$.
\end{enumerate}

For the purposes of this paper we assume that supply and demand equilibrate at a linear rate with respect to the difference between the realized labor demand and the target labor demand, and require that this relationship be non-negative.  This leads to the functional forms
\begin{equation}
    \alpha_{u,i,t} = \gamma_u \frac{\max \big\{0,  d_{i,t} - d_{i,t}^\dagger \big\}}{e_{i,t}},
    \label{eq:alpha_u}
\end{equation}
\begin{equation}
    \alpha_{v,i,t} =  \gamma_v \frac{\max \big\{0,   d_{i,t}^\dagger-d_{i,t}  \big\}}{e_{i,t}},
     \label{eq:alpha_v}
\end{equation}
where $\gamma_u$ and $\gamma_v$ are parameters that determine the speed of adjustment. They are in the interval $[0,1]$: $\gamma_u = \gamma_v = 1$ corresponds to the maximum adjustment speed\footnote{In the special case when $\delta_u = \delta_v$ it corresponds to immediate adjustment (see Eq. 19 of the Supplementary Information).} and  $\gamma_u = \gamma_v = 0$ corresponds to no adjustment at all. The $\alpha$'s are probabilities, and so must satisfy $0 \le \alpha_{u,i,t} \le 1$ and $0 \le \alpha_{v,i,t} \le 1$\footnote{Although this condition is normally satisfied automatically, there are exceptional circumstances where it would exceed the upper interval, in which case we set $\alpha = 1$.}. For the purposes of this paper we let the rates for separations and vacancies be the same, i.e. $\gamma_u = \gamma_v = \gamma$.  For a description of how we calibrated parameters and set initial conditions see the \textit{Methods} section.

All of the processes corresponding to $\delta_u$, $\delta_v$, $\alpha_{u,i,t}$ and $\alpha_{v,i,t}$ are independent.  Thus the probability that a worker in occupation $i$ is \textit{not} separated from her job is $ (1 - \delta_u)(1 - \alpha_{u,i,t})$. This means that the probability that a worker \textit{is} separated is given by
\begin{equation}
    \pi_{u,i,t} = 1-(1 - \delta_u)(1 - \alpha_{u,i,t})  = \delta_u + \alpha_{u,i,t} -  \delta_u \alpha_{u,i,t},
\label{eq:pi_u}
\end{equation}
where the negative term on the right hand side avoids counting a worker as separated twice. Similarly, for each employed worker in occupation $i$, the probability that a vacancy opens is
\begin{equation}
    \pi_{v,i,t} = \delta_v + \alpha_{v,i,t} - \delta_v \alpha_{v,i,t}.
\label{eq:pi_v}
\end{equation}

\paragraph{Automation shocks.}  We assume that automation reallocates labor demand across occupations, decreasing the number of jobs available in some professions and increasing them in others.  Since the set of occupations is fixed, we base the creation of new jobs on the thought experiment that each non-automated job reduces work hours so that the total number of jobs in the economy stays constant. This assumption is motivated by the long-run evidence that unemployment rates have no trend but hours worked have decreased \cite{ramey2009century}. We assume that the aggregate demand remains constant during the shocks (i.e., $D^{\dagger}_{t} = D_{0} = L$); automation reduces the target demand for occupations with a high automation level and correspondingly increases the target demand for occupations with a lower automation level, so that the number of jobs destroyed equals the number of jobs created (see \textit{Methods}). In the Supplementary Information we relax the assumption of fixed aggregate demand and investigate the behavior under lower and higher aggregate demand.

This completes our specification of the model.  Table S2 of the Supplementary Information 
gives a summary of the variables and parameters, the calibration procedure and Table 1 contains fitted values for all the parameters.

\subsection*{Deterministic approximation for large populations}
Although the workers and employers follow simple rules in our model, when the number of workers $L$ is large, running the computer simulation is computationally costly. This is mostly due to the number of choices unemployed workers have, the creation and closing of vacancies, and the employers' selection of workers at each time step. However, when $L$ is large, we can take advantage of the law of large numbers and multivariate Taylor expansions to approximate the system's behavior in terms of expected values. This provides a good approximation for most purposes and is faster to simulate and easier to analyze, which is very useful for exploring the parameter space when calibrating the model. For brevity, we keep the approximations and derivations in the Supplementary Information.

We compute expectations for Eqs.~(\ref{eq:e_t_exact} - \ref{eq:v_t_exact}) in the limit of a large number of agents and conditional on the state of the system at the previous time step. To keep the notation compact, we often denote expected values by a bar above the variable, e.g., 
$$
 \bar{u}_{i, t+1} \equiv E \left[u_{i, t+1}| \mathbf{u}_{i,t}, \mathbf{v}_{i,t}, \mathbf{e}_{i,t}\right].
 $$
We reduce the master equations to a $3n$ dimensional deterministic dynamical system given by Eqs.~(\ref{eq:e_t} - \ref{eq:v_t}).
\begin{eqnarray}
 \label{eq:e_t}
\bar{e}_{i,t+1} & = & \bar{e}_{i,t}  - \  \underbrace{ \Bigg( \delta_u  \bar{e}_{i,t} +  (1 - \delta_u) \gamma_u \max \big\{0,  \bar{d}_{i,t} - d_{i,t}^\dagger \big\} \Bigg)}_\textrm{separated workers} \ + \ \underbrace{\sum_j  \bar{f}_{ji, t+1}}_\textrm{hired workers} ,\\
 \label{eq:u_t}
\bar{u}_{i,t+1} &  =  & \bar{u}_{i,t} + \underbrace{ \Bigg(   \delta_u  \bar{e}_{i,t} +  (1 - \delta_u) \gamma_u \max \big\{0,  \bar{d}_{i,t} - d_{i,t}^\dagger \big\}  \Bigg)}_\textrm{separated workers} \ - \ \underbrace{\sum_j \bar{f}_{ij, t+1}}_\textrm{transitioning workers} ,\\
\label{eq:v_t}
\bar{v}_{i,t+1} & = & \bar{v}_{i,t} +  \underbrace{ \Bigg( \delta_v \bar{e}_{i,t} +  (1 - \delta_v) \gamma_v \max \big\{0, d_{i,t}^\dagger - \bar{d}_{i,t} \big\} \Bigg)}_\textrm{opened vacancies} \ - \  \underbrace{\sum_j \bar{f}_{ji, t+1}}_\textrm{hired workers} ,
 \end{eqnarray}

As we show in the Supplementary Information, we can write $\bar{f}_{ij,t+1}$ in terms of the adjacency matrix and the expected values of the state variables as
\begin{equation}
\bar{f}_{ij,t+1} =  \frac{\bar{u}_{i,t} \bar{v}^2_{j, t} A_{ij} (1 - e^{-\bar{s}_{j,t+1}/\bar{v}_{j,t}})}{\bar{s}_{j,t+1}\sum_k \bar{v}_{k,t} A_{ik}} , 
\label{eq:f_full}
\end{equation}
where
\begin{equation}
\bar{s}_{j,t+1} = \sum_i \frac{\bar{u}_{i,t} \bar{v}_{j, t} A_{ij}}{\sum_k \bar{v}_{k,t} A_{ik}}.
\end{equation}
The relative error of our approximation is 
$$\left| \frac{ E[f_{ij,t+1}| \mathbf{u}_t, \mathbf{v}_t ; A] - \bar{f}_{ij,t+1}}{E[f_{ij,t+1}| \mathbf{u}_t, \mathbf{v}_t ; A] }\right| < \frac{c}{L + c},$$
where $c$ is a constant, here $E[f_{ij,t+1}| \mathbf{u}_t, \mathbf{v}_t ; A]$ denotes the expected value and $\bar{f}_{ij,t+1}$ the expected value in the limit of a large number of agents. Using Eqs \ref{eq:u_t} and \ref{eq:f_full} we compute long-term unemployment (see \textit{Methods} for details).

Eqs. (\ref{eq:e_t} -- \ref{eq:v_t}) allow us to study the system, in this case, the U.S. labor market, in a tractable and timely manner. Given a set of time series for the target labor demand $d^\dagger_{i,t}$ and a set of initial conditions, Eqs. (\ref{eq:e_t} -- \ref{eq:v_t}) determine the expected employment, unemployment, and vacancies as a function of time. All of our results are based on the U.S. occupational mobility network, which classifies jobs into 464 occupational categories. In the Supplementary Information, we show that the deterministic approximation derived above provides a good approximation when using the U.S. occupational mobility network as long as each occupation has a target demand of at least $50$ workers. If we assume a labor force of 1.5 million, almost all occupations satisfy this. Thus for most occupations, the deterministic approximation is valid for any labor pool bigger than that of a medium-sized city. Our results can thus be thought of as applying to a city of at least this size, under the assumption of perfect job mobility within the city.  We will also assume that the occupational mobility network $A$ for this ``typical city" is that of the U.S. as a whole.  Because we will later calibrate the model based on national occupational unemployment levels, we implicitly assume that these reflect the national average.

\paragraph{Steady-state}\label{sec:steady-state}
Before proceeding to analyze the U.S. labor market under a changing demand for labor, we note that, when the target labor demand is constant $d_i^\dagger$, there exists a computable steady-state value for the number of employed and unemployed workers and vacancies in each occupation. Except for the simple case of a complete network, with $A_{ij} = 1/n$, we cannot derive a closed-form solution for the occupational unemployment. Nonetheless, we can solve equations numerically to find the solution. In the Supplementary Information we also show that the steady-state values depend on the network structure as well as the target labor demand. Thus the network structure, and the distribution of labor demand on it, can substantially influence the steady-state unemployment at both the occupational and the aggregate level.


\subsection*{The Beveridge curve}
\begin{figure}
\centering
\includegraphics[width=0.9\linewidth]{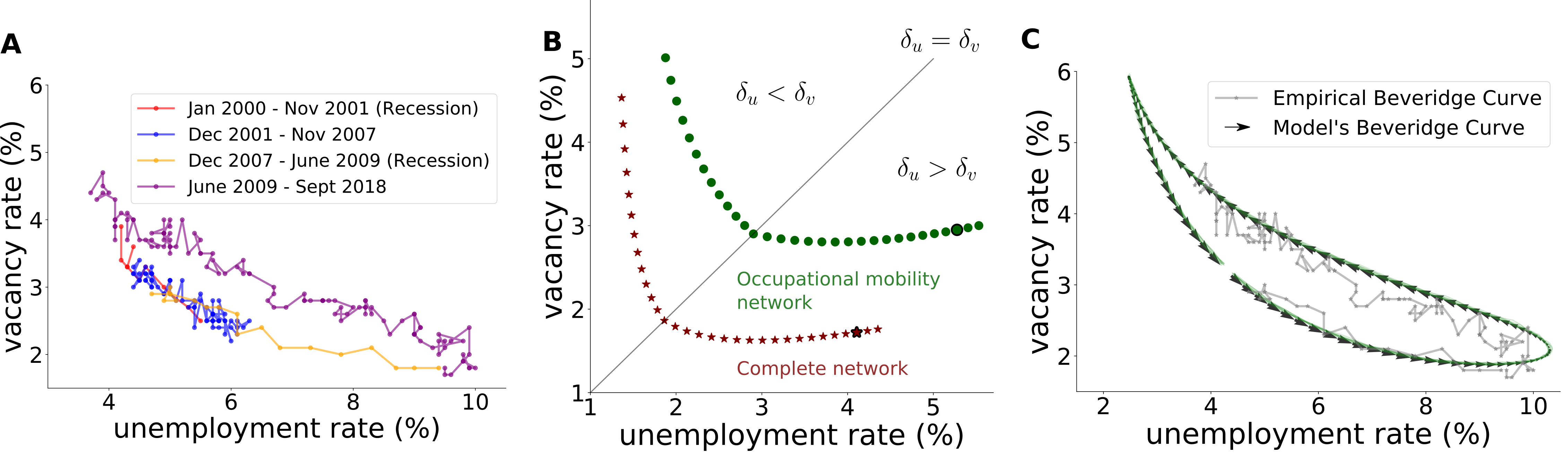}
\caption{\textbf{The Beveridge curve.} In each panel we plot the unemployment and vacancy rate. (\textbf{A}) The historical Beveridge curve for the United States, 2000-2018. Different periods are highlighted with different colors. (\textbf{B}) Movement of the Beveridge curve due to changes in labor market frictions. In particular, we plot the difference between a complete network, with no skill mismatch frictions, vs. the empirical occupational mobility network. Each dot corresponds to the steady-state unemployment and vacancy rate for different values of $\delta_u$ and $\delta_v$. The highlighted points correspond to the unemployment and vacancy rate of the model using the calibrated parameters. (\textbf{C}) The Beveridge curve generated by our model. The parameters of the model are calibrated to match the empirical Beveridge curve between December 2007 and December 2018. The dashed lines correspond to the deterministic approximation of Eqs.~(\ref{eq:e_t} - \ref{eq:v_t}) and solid green lines to the full stochastic model simulation of Eqs.~(\ref{eq:e_t_exact} - \ref{eq:v_t_exact}).  The transparent grey line shows the empirical Beveridge curve between December 2007 and December 2018.}
\label{fig:Beveridge_Curve}
\end{figure}

The Beveridge curve is one of the most well-known macroeconomic stylized facts \cite{diamond2015shifts, beveridge2014full}. It states the relation between vacancies and unemployment:  When more vacancies open, unemployment goes down.  
The intuition is that when there are many vacancies, unemployed workers get a job faster, so the unemployment rate is low. Similarly, when there are few vacancies, unemployed workers are less likely to find jobs, so the unemployment rate is high. 
In panel A of Fig. \ref{fig:Beveridge_Curve} we plot the Beveridge curve for the USA between January 2001 and September 2018.

There are three important features of the Beveridge curve: (i) The curve can shift away or toward the origin \cite{diamond1982aggregate}. For example, after the 2009 financial crisis, the Beveridge curve shifted away from the origin, with unemployment increasing for all vacancy rates. (ii) During recession periods, unemployment and vacancy rates move downward along the curve, and during recovery periods, the unemployment and vacancy rates move upward along the curve. The recession from December 2007 to June 2009 and the recovery period from 2009 onwards are a good example of this feature (see panel A of Fig. \ref{fig:Beveridge_Curve}). (iii) Historically the Beveridge curve has (almost) always shifted outwards after recessions \cite{diamond2015shifts}, i.e., the curve cycles counter-clockwise. In other words, for the same vacancy rate, the unemployment rate has been larger during recoveries than during recessions. As we show later, our model reproduces these three features. 

We calibrate the parameters of our model using the Beveridge curve. To do so, we impose a simulated business cycle. For simplicity, we assume that the aggregate target labor demand $D_t$ oscillates according to a sine wave. We then calibrate the amplitude of the sine wave and the parameters $\delta_u$, $\delta_v$, and $\tau$ to match the empirical Beveridge curve during the most recent U.S. business cycle, from 2008 to 2018.  We set the initial target labor demand of each occupation equal to the observed average employment in 2016 (see calibration details in the Supplementary Information). 

Our model reproduces the three mentioned features of the Beveridge curve: First, we show that structural changes, such as a decrease in the efficiency of worker-vacancy matching, cause the Beveridge curve to shift with respect to the origin. To demonstrate this, we now hold the target aggregate demand $D_t$ constant, and instead, vary the structure of the network by replacing the empirical network $A$ by a complete network $A_{ij} = 1/n$, in which each node is linked to every other node with equal weights. This corresponds to the null hypothesis of no skill restrictions.  We do this for different values of $\delta_u$ and $\delta_v$ and trace the {\it steady-state} behavior in Fig. \ref{fig:Beveridge_Curve}B. 
 As expected, when we remove the network structure, the Beveridge curve shifts downwards towards the origin. When we consider parameters calibrated to actual data (highlighted with a bold border), removing the network structure corresponds to an increase in unemployment from $4.1\%$ to $5.3\%$. This effect is substantial, representing more than a $25\%$ increase. Second, our model reproduces the dynamics of the Beveridge curve over business cycles.  As we show in Fig. \ref{fig:Beveridge_Curve}C, unemployment and vacancy rates move downward along the curve during recessions, and upward along the curve during recovery periods. Third, the Beveridge curve of our model cycles counter-clockwise, shifting away from the origin after recession periods. 

The first two features have been explained by several models (see \cite{elsby2015beveridge} for a review). Most prominently, the Diamond-Mortensen-Pissarides model predicts that structural changes in the labor market cause the Beveridge curve to shift. For example, in their model, an increase in skill mismatches would decrease the number of worker-vacancy matches, shifting the Beveridge curve to the right. The first feature was also explained by Axtell et al. \cite{axtell2019frictional} who, similar to us but using synthetic networks instead of empirical ones, showed that changes in the network structure could shift the Beveridge curve closer or further away to the origin. The Diamond-Mortensen-Pissarides model also explains that during booms of the business cycle, more vacancies open and unemployment decreases, while in recessions, fewer vacancies open, and more workers are separated. Thus, they correctly predict the second feature: points on the upper left of the curve correspond to booms and points on the lower right to recessions \cite{pissarides2011equilibrium,diamond1982aggregate}.
The third feature (the counter-clockwise cyclicality of the Beveridge curve) is yet to be fully understood. The standard interpretation is that the outwards shifts of the Beveridge curve correspond to a deterioration in the matching/hiring process in the economy \cite{diamond2015shifts}. However, some equilibrium models \cite{mortensen1999equilibrium,pissarides1985short,sniekers2018persistence} argue that this phenomenon is independent of structural change and due to the business cycle dynamics. 
In these models, vacancies can adjust immediately following the decisions of the producers, unlike unemployment, which decreases only due to a successful job matchings. This difference in the flexibility of the variables causes vacancies to increase faster than employment during the recovery, and thus the Beveridge curve cycles counter-clockwise \cite{pissarides1985short,sniekers2018persistence}. 

Our model supports the hypothesis that the counter-clockwise cyclicality of the Beveridge curve results naturally from the business cycle dynamics. As we show in Fig. \ref{fig:Beveridge_Curve}C, when we use the calibrated parameters, the Beveridge curve cycles in a counter-clockwise direction even though we assume no structural changes. Both the choice of parameters and the type of network used in the model influence the Beveridge curve's position, direction of cycle, and area it encloses (see Supplementary Information for examples). 


Undertaking a detailed assessment of the exact behavior of the Beveridge curve under different parameter choices is out of scope for this paper. We do, however, briefly discuss how the direction in which the Beveridge curve cycles is influenced by the state-independent rates at which workers are separated ($\delta_u$) and vacancies open ($\delta_v$).


We run a number of numerical exercises varying $\delta_u$ and $\delta_v$ while keeping all other parameters fixed. Starting from the calibrated values $\delta_u = 0.016$ and $\delta_v = 0.012$, we gradually decrease $\delta_u$ and increase $\delta_v$ until $\delta_u = 0.012$ and $\delta_v = 0.016$. During these gradual changes, we find that the Beveridge curve first reduces its enclosed area, then changes its cycling direction from counter-clockwise to clockwise, and finally increases its enclosed area (see figures in the Supplementary Information). This behavior is present for both the occupational mobility network and the complete network's Beveridge curve. 

We also find that as $\delta_u$ takes on greater values than $\delta_v$, the Beveridge curve tends to cycle in a counter-clockwise direction\footnote{The difference between $\delta_u$ and $\delta_v$ needed for the Beveridge curve to cycle counter-clockwise depends on the network.}; i.e., the recovery of unemployment is slow, meaning that for the same vacancy rate, the unemployment rate will be higher during recoveries than during recessions.
In summary, if the state-independent rate at which workers are separated is greater than the rate at which vacancies open (i.e., $\delta_u > \delta_v$), there is a tendency for workers to become unemployed faster than vacancies open. As such, it is not surprising to find a slower unemployment recovery under these parameter settings. 

\subsection*{The impact of automation on employment}
We now use the model to assess the impact of automation shocks on employment. We study two automation scenarios, one based on a study by Frey and Osborne \cite{frey2017future} and the other based on a study of Brynjolfsson et al. \cite{brynjolfsson2018SML}. We refer to these automation scenarios as the {\it Frey and Osborne shock} and as the {\it Brynjolfsson et al. shock}. For brevity, we show the figures for the Brynjolfsson et al. shock in the Supplementary Information. 
 
\paragraph{Estimates of the automation shock}\label{subsubsec:results_automationhyp}
Frey and Osborne estimated the probability that each of 702 occupations in the O*NET 6-digit classification system could be computerized soon \cite{frey2017future}. To do this, they gave experts a description of tasks performed by workers in a restricted sample of 70 occupations and asked them whether the occupations could be automated within the next two decades. Based on the experts' answers and using nine O*NET variables that describe occupations as inputs, they trained a supervised machine learning algorithm and estimated what they called the {\it probability of computerization} for the remaining occupations.  They found that approximately half of the jobs in the U.S. would be at risk for some degree of automation.  

This study, as well as the Brynjolfsson et al. study (see Supplementary Information),
estimate the probability that an occupation will be \emph{technically automatable}. This is not the probability that an occupation will be automated, which also depends on cost, institutions, etc., and it is not an estimate of the share of jobs in an occupation that will be automated.  
Nonetheless, for simplicity we interpret these as automation levels, directly determining the share of jobs in an occupation that will be automated. We map the 6 and 8 digit O-NET classifications used in these studies into the the U.S. occupational mobility network (which is based on the 4 digit American Community Survey classification) using the 2016 National Employment Matrix Crosswalk (see \cite{mealy2018}).

\begin{figure}
\centering
\includegraphics[width=0.6\linewidth]{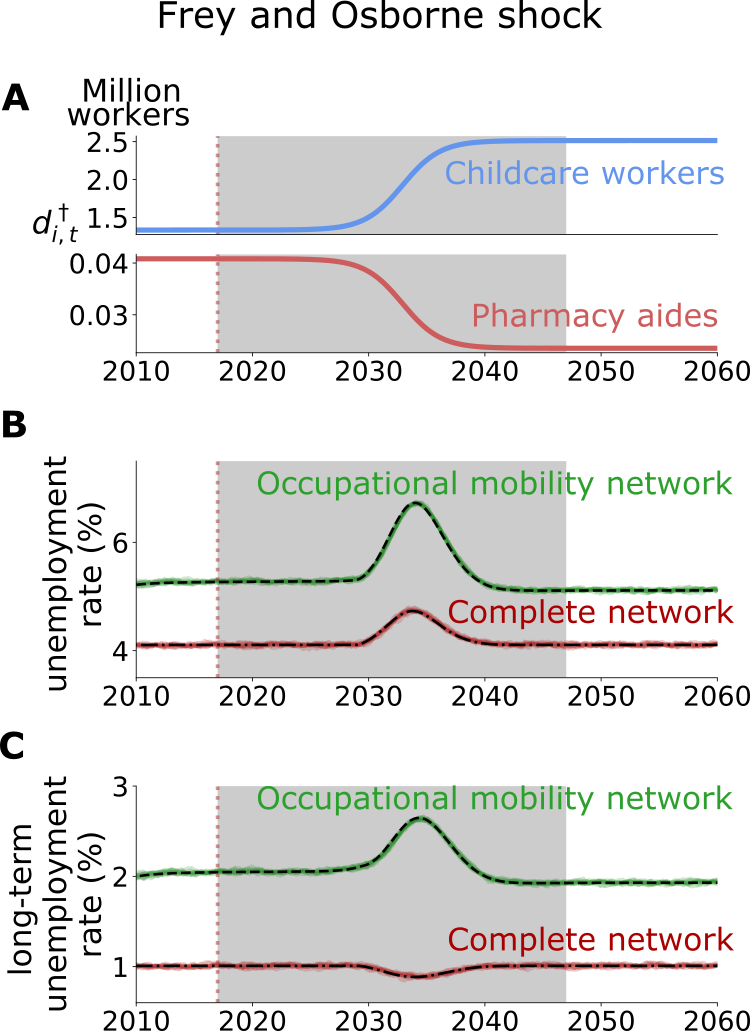}
\caption{\textbf{Aggregate labor market outcomes under the Frey and Osborne shock.} The grey area denotes the 30 years during which the automation shock takes place. Panel (\textbf{A}) shows the evolution of the target labor demand for two example occupations. The occupation colored in blue has a low automation level and the occupation colored in red has a high level. Because of its heterogeneity across occupations, the Frey and Osborne shock implies a large change in the target labor demand of most occupations. Panel (\textbf{B}) shows the unemployment rate as a function of time. Dashed lines are our approximations of the expected value (solved numerically) and the solid lines are 10 simulations with 1.5 M agents. Panel (\textbf{C}) shows the long-term unemployment rate as a function of time. As before, dashed lines correspond to the deterministic approximation of Eqs.~(\ref{eq:e_t} - \ref{eq:v_t}) and solid lines to the full stochastic model simulation of Eqs.~(\ref{eq:e_t_exact} - \ref{eq:v_t_exact}).}
\label{fig:automation_shock}
\end{figure}

\paragraph{Introducing automation shocks}
Before the automation shock, we assume the system is in a steady-state where the target demand $d^\dagger_{i,0}$ matches the employment distribution in $2016$ (see Material and Methods). We then introduce an automation shock by making the target demand $d_{i,t}^\dagger$ follow a sigmoid function, which begins at $d^\dagger_{i,0}$ and converges to the post-automation target demand (see top panel of Fig. \ref{fig:automation_shock} for examples). We choose the adoption rate so that the total shock is spread across a $30$ year period, though most of the change happens within about $10$ years.  See \emph{Methods} for details, and the Supplementary Information, where we show that the results are fairly robust for reasonable adoption rates.  


\paragraph{Aggregate level outcomes.} 
As seen in Fig. \ref{fig:automation_shock}, even though the aggregate target demand is held constant, the Frey and Osborne shock increases both the aggregate unemployment rate and the aggregate long-term unemployment rate during the period of automation. This increase is caused by the substantial reallocation of labor demand across occupations (see panel A for an example of how the target demand changes at the occupation level).

We compare the behavior with the empirical occupational mobility network to the hypothetical behavior assuming a complete network, in which any worker can transition equally well to any occupation. We use  the same parameters for both networks (see calibrated parameter values in Table \ref{tab:parameter_values} in Methods section).  The aggregate unemployment rate is initially about $5.3\%$ for the empirical network and $4.1\%$ for the complete network. When we apply the Frey and Osborne shock, the aggregate unemployment rate for the empirical network rises to $6.7\%$ at its peak and then decays. In contrast, for the complete network the aggregate unemployment rises to only $4.7\%$ before it decays. Thus the total change in unemployment with the empirical network is more than a factor of two larger, demonstrating the importance of the network structure. 

We also study the behavior of the long-term unemployment as a function of time. For the same parameter choices, the long-term unemployment rate for the empirical network is about $2.0\%$, substantially smaller than the (short-term) unemployment.  
When we apply the Frey and Osborne shock the aggregate long-term unemployment rate rises to $2.6\%$ at its peak and then decays.  The relative change from the initial value to the peak value is about $29\%$, which is similar to the relative change of $27\%$ for the unemployment rate.  However, the behavior for the complete network is quite different: First, the initial level of long-term unemployment for the complete network is only $1.0\%$, more than a factor of two smaller than for the empirical network. Second, when we apply the shock, long-term unemployment for the complete network remains nearly flat.

Another surprising result is that the steady-state value of the aggregate unemployment shifts after the shock.  The aggregate unemployment rate changes from $5.27\%$ to $5.11\%$, for a net change of about $-0.15\%$.  While this is small, bear in mind that we have kept the both the total aggregate target demand and all the parameters of the model constant.  This is consistent with our result that the steady-state explicitly depends on the network structure and the target demand in each occupation (see Supplementary Information Eqs. 22--24 for details). The fact that we see this shift when we change the target demand demonstrates the key role that the network structure plays in determining the steady-state as well as the transient behavior. Note that there is no noticeable shift in the steady-state for the complete network. 

We conjecture that the Frey and Osborne shock causes such persistent effects since automation levels of neighboring occupations tend to be similar. This has two effects:  It means that there are some regions of the network where workers easily find new jobs, and others where workers get trapped because there are no good alternatives, causing a substantial boost to long-term unemployment. The shift in the steady-state occurs because the post-automation distribution of the target labor demand across occupations is more concentrated on fewer occupations that are more densely connected between each other, reducing worker-vacancy matching frictions. We test our conjecture by creating a surrogate Frey and Osborne shock that randomizes the distribution of automation levels of occupations across the network and find supporting results (see Supplementary Information for details).

These results demonstrate how the structure of the occupational mobility network can cause substantial and long-lasting dislocations of the labor force.  When the automation shock alters the target demand for labor, old jobs are closed in some occupations, and new jobs are opened in others.  Under the Frey and Osborne shock, which has a strong network structure, we see a sizeable transient effect that causes a substantial rise in aggregate unemployment over the decade during which the dislocation takes place.  This is also felt in aggregate long-term unemployment and even causes a permanent shift in aggregate unemployment.    

Finally, in the Supplementary Information, we explore what happens when we relax the assumption that the aggregate target labor demand remains constant.  

\paragraph{Occupation level outcomes.}

We now show how automation affects the occupation-specific unemployment rates, where the network plays a crucial role. To avoid problems with small denominators, and to ensure that each unemployed worker contributes equally to the average unemployment rate during the automation period, we measure the \textit{average unemployment rate} and \textit{average long-term unemployment rate} during the shock as 
\[
u_{i,\text{average}}(T)= \frac{\sum_{t\in T} u_{i,t}}{\sum_{t\in T} (u_{i,t} + e_{i, t})},
\]
and
\[
u_{i,\text{average}}^{(\geq \tau)}(T)= \frac{\sum_{t\in T} u^{(\geq \tau)}_{i,t}}{\sum_{t\in T}, (u_{i,t} + e_{i, t})},
\]
where $T$ is the set of time steps that correspond to the automation shock.  (We discuss an alternative way of defining the average unemployment rate in the Supplementary Information). For simplicity, from here onward, we refer to the average unemployment rate and the average long-term unemployment rate during the automation period simply as the unemployment rate and the long-term unemployment rate.

\begin{figure}
\centering
\includegraphics[width=0.75\linewidth]{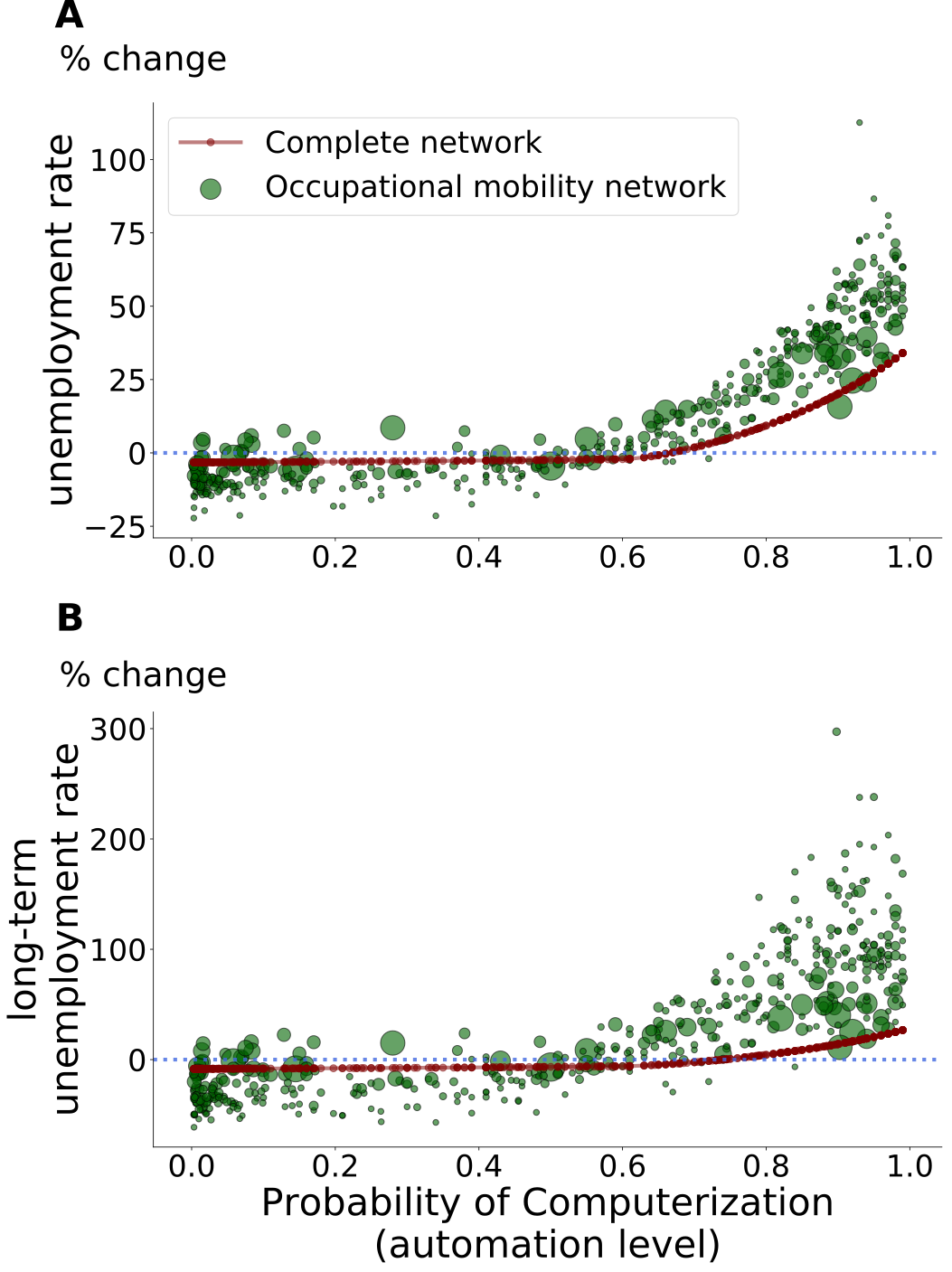}
\caption{\textbf{Impact of the Frey and Osborne shock on unemployment and long-term unemployment at the occupation level.} The green dots are for the occupational mobility network and the red dots are for the complete network. The size of the green dots is proportional to the employment of the occupation they represent. Panel \textbf{(A)} shows the percentage change in the unemployment rate vs the automation level for each occupation, while panel \textbf{(B)} shows the same thing for the long-term unemployment rate. The scatter in the results demonstrates that, due to network effects, the automation level only partially explains occupational unemployment.}
\label{fig:lt_unemp_10 years_FO}
\end{figure}

 In Fig. \ref{fig:lt_unemp_10 years_FO} we compare the percentage changes in unemployment and long-term unemployment with the automation level of each occupation.  To highlight the role of the network, we do this both for the occupational mobility network, which includes market frictions due to skill mismatch, and for the complete network, where workers can apply to any job vacancy regardless of their occupation. For the complete network, the automation level of an occupation uniquely determines the impact of automation, i.e., occupations with the same automation level have the same percentage change in their unemployment and long-term unemployment rates.  In contrast, for the occupational mobility network, due to network effects, there is considerable scatter around the mean behavior -- unlike the complete network, the automation level is {\it not} a perfect predictor of the occupation-level outcome. The scatter is substantial both for the Frey and Osborne shock and for the Brynjolfsson et al. shock.  (See also Tables S2 and S3 of the Supplementary Information).
 
To make the size of these effects clear it is useful to highlight some specific cases. Both dispatchers and pharmacy aides have a high probability of computerization of $0.72$, but the automation shock causes a $19\%$ increase in the dispatchers' long-term unemployment, while the pharmacy aides' long-term unemployment {\it decreases} by roughly by the same ratio. 
Some occupations experience the opposite change that one would expect. Statistical technicians and pharmacy aides are likely to be automated (with a probability of computerization above $0.6$) while childcare workers and electricians are not (with probability of computerization below $0.1$). However, statistical technicians and pharmacy aides {\it decrease} their long-term unemployment, while childcare workers and electricians increase theirs. This is due to the fact that it is relatively easy for statistical technicians and pharmacy aides to transfer to jobs in other occupations with increasing demand. In contrast, it is easier for others to transfer to childcare workers or electricians, thereby increasing the supply of workers relative to the demand. This illustrates the importance of network effects.


\paragraph{Brynjolfsson et al. automation shock}
Brynjolfsson et al. estimated the \textit{suitability for machine learning} of occupations, which we use as an alternative hypothesis for the automation shock (see Supplementary Information for details). 
Unlike the Frey and Osborne shock, the Brynjolfsson et al. shock causes no noticeable change in the aggregate unemployment rates. This difference is caused by the different distributions of the two shocks. The Frey and Osborne shock is very heterogeneous across occupations, affecting some occupations a great deal and others very little, so that the changes in target demand at the occupation level are substantial (see Fig. \ref{fig:network_fo} panel A). In contrast, the Brynjolfsson et al. shock affects most occupations similarly, so that the changes in the target demand are lower and the network effects are small (see Fig. S3 panel A in the Supplementary Information). However, during the Brynjolfsson et al. shock, we still observe the network effects at the occupation level. The change in the long-term unemployment and unemployment varies substantially for occupations with similar suitability for machine learning. (see Fig. S3 panel A in the Supplementary Information).

\section*{Discussion}
This paper develops a new out-of equilibrium model of the labor market and applies it to analyze the impact of automation on unemployment. At the occupation level, we show that employment impacts for workers are likely to depend not only on the automatability of their current occupation, but also the alternative occupations that they can transition into. At the macro level, our model reproduces the dynamics of the Beveridge curve. In contrast to standard models \cite{diamond1982aggregate,axtell2019frictional}, we find that the Beveridge curve can shift outwards after recessions without introducing structural changes. This finding supports the hypothesis that the  counter-clockwise cyclicality can be caused solely by business cycle dynamics alone \cite{kohlbrecher2016business,sniekers2018persistence}. 

Similar to previous studies \cite{axtell2019frictional} we find that the occupational mobility network structure affects the unemployment rate. However, here we go further by quantifying the labor market frictions imposed by the empirical network: these frictions can account for up to $25\%$ of the steady-state unemployment rate. We also find that the distribution of labor demand across occupations in the network can affect the steady-state unemployment, as we demonstrated in the Frey and Osborne automation shock. Most importantly, in studying the transient period associated with an automation shock, we find that even when the total number of jobs remains constant, automation can increase long-term unemployment due to the mismatch between unemployed workers and job vacancies. 

Our work complements previous efforts that have studied automation and job displacement  based on the {\it task approach\/}  \cite{autor2013task,acemoglu2011skills,acemoglu2018artificial}, but provides a networks perspective on job transitions that goes beyond classifying workers into low, middle and high skill categories. Our paper is also closely related to work that has used networks to study the effects of labor market frictions\cite{guerrero2013employment,axtell2016network, neffke2018mobility} or the propagation of economic shocks\cite{nimczik2017job,diodato2014resilience,dworkin2019network}. However, as these studies assume independence between workers (meaning that worker-worker displacement is ignored), focus on either equilibrium or steady-state dynamics, and/or do not yield economic variables such as unemployment and long-term unemployment, we believe our work makes an important and unique contribution. 

Our findings are also particularly relevant for the macroeconomic literature on the Beveridge curve. Studies based on search theory and networks have argued that the shifts of the Beveridge curve can be caused by structural changes \cite{petrongolo2001looking,rogerson2005search,axtell2019frictional}. Meanwhile, other studies suggest that these shifts are part of the counter-clockwise cyclicality of the Beveridge curve, which results from business cycles dynamics \cite{kohlbrecher2016business,sniekers2018persistence}. Our work supports this latter hypothesis: business cycles alone are enough to cause the Beveridge curve to cycle counter-clockwise. However, we also show that structural changes, such as changes in the network structure, cause shifts in the Beveridge curve.

\paragraph{Policy implications and future work} 
Several studies have focused solely on the automatability of occupations when assessing the outlook of workers. We propose a more complete view, by considering not only the automatability of occupations but also workers' possibilities for transitioning into occupations with open vacancies. In some cases, this perspective yields different - and counter-intuitive results, where some occupations which are expected to be at high risk of automation (such as statistical technicians) could have more promising future employment prospects, while other ‘safer’ occupations (such as childcare workers) could see higher rates of long-term unemployment. 

Such insights are likely to be important in helping today’s workforce best prepare for tomorrow’s labor market challenges and opportunities. Our model could also be particularly useful in helping policymakers target employment assistance packages and skill development programs to workers who are more likely to face longer periods of unemployment. And while this particular paper has focused on labor market shocks relating to automation, our model is quite general, and could also be adapted to analyze impacts arising from changes in labor demand relating to offshoring\cite{antras2006offshoring,blinder2013alternative} or the transition towards the green economy\cite{greenstone2002impacts,morgenstern2002jobs,walker2011environmental}. 

There is, of course, plenty of room for additional research. One can further explore how the model's parameters change the behavior of the Beveridge curve and improve the calibration by using data on multiple business cycles. We have not yet considered the role of geography \cite{diodato2014resilience,neffke2008relatedness} or the feedback effects from the production network \cite{jackson2019automation}, and in contrast to several labor market models \cite{diamond1982aggregate,rogerson2005search}, we assume inelastic labor demand and neglect wage dynamics. This is a simplifying assumption that allows us to focus on labor market frictions due to worker-vacancy mismatches. Adding wages into the model would be an important step forward. However, doing so in a realistic manner requires vacancy data at the occupation level. While such data is so far not publicly available, work is underway to prioritize data collection efforts to facilitate labor market research \cite{frank2019toward}. 

\section*{Materials and Methods}

\subsection*{Labor reallocation due to labor automation}
While it is clear that automation will replace a number of workers, this is an old process that has so far not caused persistent large unemployment rates \cite{frank2019toward}. Instead, the average length of the work week has declined substantially \cite{ramey2009century} and work has shifted to new occupations \cite{lin2011technological}. Thus, we assume that automation will lead to a post-shock reallocation of labor demand, with some occupations increasing and others decreasing their labor demand. 

Our model requires us to specify the post-shock reallocated demand $\mathbf{d}^\dagger$, which is the value to which the target labor demand converges  after the shock. As we explain here, we use the the probability of computerization scores of Frey and Osborne\cite{frey2017future} and the suitability for machine learning scores of Brynjolfsson et al.\cite{brynjolfsson2018SML} to set the post-shock reallocated demand. The time at which the target demand converges to the post-shock reallocation demand if $t^*$.

First, we set the level of automation in each occupation, which is bounded by 0 and 1, equal to the  probability of computerization or to the normalized suitability for machine learning scores\footnote{We divide the score by 5, which is the maximum possible score, to normalize the scores between 0 and 1} depending on the shock. 
We assume that the level of automation is the fraction of total hours worked in an occupation that are no longer needed post-shock. Furthermore, working hours are reduced for all workers in the economy, so that the total number of jobs stays constant. We denote the labor force, which is the number of workers, by $L$ and assume that it remains constant. Let $x_0$ be the current number of hours of labor for the average worker in a given period of time (say a year). The hours of work each occupation demands is given by the components of the vector
\[
 \mathbf{h}_0 = x_0 \mathbf{e}_0.
\]
Letting $\mathbf{p}$ be the vector with the automation level of each occupation, the new number of hours of work $\mathbf{h}_{t^*}$ after automation is
\[
 \mathbf{h}_{t^*} = \mathbf{h}_0 \odot ( \mathbf{1} - \mathbf{p}),
\]
where $\odot$ denotes the element-wise multiplication of vectors and $\mathbf{1}$ the vector of ones.  We split the aggregate hours of work equally among workers, thus the number of hours of work per week is
\[
 x_{t^{*}} = \frac{\sum_i^n h_{i,t^{*}} }{L}.
\]
Finally, assuming that automation has no impact on the aggregate labor demand unemployment, we split the hours of labor demanded by occupations equally among workers.
\begin{equation}
  \mathbf{d}_{t^*}^\dagger  \equiv  \mathbf{d}^\dagger  = \mathbf{h}_{t^*} \frac{1}{x_{t^*}}.
\end{equation}
where $t^*$ is the time at which the target labor demand reached the post-shock target. In the Supplementary Information we explore the behavior under an aggregate increase or decrease in the number of jobs.
 
\subsection*{Formulating a time dependent automation shock}
We follow the innovation literature, which suggests that the adoption of technologies follows a sigmoid function or S-curve over time \cite{stoneman2001economics}. Frey and Osborne say that their estimates are over ``some unspecified number of years, perhaps a decade or two.'' \cite{frey2017future}. 
We assume that the automation happens within $30$ years, but mostly happens within $10$ years, and explore different alternatives in the Supplementary Information. 

We assume that the target demand initial value is the steady-state demand and over time it reaches the post-shock reallocated demand $\mathbf{d}^\dagger$. Within $15$ years, the target demand is at the mid-point between the initial steady-state demand and the post-shock reallocated demand. 
We use a sigmoid function for the target demand
\begin{equation}
  d_{i, t}^\dagger = 
  \begin{cases} 
    d_{i, 0} & \text{if } t < t_s  \\
   d_{i, 0} + \frac{d_i^\dagger - d_{i,0}}{1 + e^{k(t - t_0)}}      & \text{if } t \geq t_s.
  \end{cases}
\end{equation}
where $t_s$ is the time at which the automation shock starts and $t_0$ is 15 years after $t_s$. Furthermore $k=0.79$, which guarantees that the target demand equals the post-shock reallocate demand up to a $0.0001$ tolerance.

Before introducing the automation shock, we first initialize the model so that it converges to the steady-state unemployment rate and to the employment distribution of occupations of $2016$. After it reaches the steady-state, we introduce the target demand $d_{i,t}^\dagger$ as explained above. 
In the Supplementary Information we demonstrate the robustness of the results under variations in the time span of the automation shock.

\subsection*{Building the occupational mobility network}\label{subsec:network}
Following Mealy et al., we construct the occupational mobility network using empirical data on occupational transitions \cite{mealy2018}. The classification is based on the 4-digit occupation codes, which yields $464$ distinct occupations.  We used monthly panel data from the US Current Population Survey (CPS) to count the number of workers $T_{ij}$ who transitioned from occupation $i$ to occupation $j$ during the period from January 2010 to January 2017.  Letting $T_i = \sum_j T_{ij}$, we assume that if a worker changes occupation, the probability of transitioning from occupation $i$ to occupation $j$ is
\begin{equation}
 P_{ij} = \frac{T_{ij}}{T_i}.
\end{equation}
For simplicity, we assume that the probability that a worker who changes jobs remains in the same occupation is constant across occupations.
Letting $r$ be the probability that a worker who changes jobs stays in the same occupation, we write the adjacency matrix of the occupational mobility network in the form 
\begin{equation}
 A_{ij}= 
  \begin{cases} 
   r & \text{if } i = j,  \\
   (1 - r)P_{ij}      & \text{if } i \neq j.
  \end{cases}
\end{equation}
We estimate $r$ based on the annual occupational mobility rate, which is the percentage of workers that switch occupations within a year \cite{groes2014u}. Specifically, we calibrate $r$ to match the number of workers that annually change occupations in a year in the model with the empirical data. 

While the empirical mobility network allows us to calibrate the heterogeneity in occupational mobility at a detailed level, the concepts are different; the relative preference with which a worker from occupation $i$ applies to a job vacancy in $j$ is different from the probability that a worker from occupation $i$, who is switching jobs, transitions to occupation $j$. However, the former is not directly observable from data. To overcome this issue, we use the occupational mobility network as indicative of the preference with which a worker from occupation $i$ applies to a job vacancy in $j$. A caveat is that since the odds of a worker being hired do not uniquely depend on the preference with which workers apply to job vacancies, the transitions of workers observed in our model do not perfectly match the empirically observed transitions. Though the matching between the transitions in our model and the empirical network is not perfect, they are significantly similar -- the Pearson correlation between them is $0.97$.

\subsection*{Calibration}
The calibrated parameters of the model are shown in Table \ref{tab:parameter_values}. For brevity, we describe the calibration methods in the Supplementary Information.

\begin{table}[ht!]
\centering
\caption{Calibrated parameter values}
\begin{tabular}{p{2.051cm}|p{2.1cm}|p{5.5cm}}
\textbf{Parameter}  & \textbf{Value}   & \textbf{Description}   \\
\hline
$\delta_u$            & 0.0160  & Rate at which employed workers are separated due to the spontaneous process.  \\
$\delta_v$            & 0.0120 & Rate at which employed vacancies are opened due to the spontaneous process. \\
$\gamma$            & 0.160  & Speed at which the realized demand adjust towards the target demand by separating workers or opening vacancies. \\
$\Delta t$            & 6.75 &  Duration of a time step in units of weeks.  \\
$r$                 & 0.55 & Probability that a worker stays in the same occupation  \\
\end{tabular}
\label{tab:parameter_values}
\end{table}

\bibliography{bibliography}

\bibliographystyle{plain}

\section*{Acknowledgments}
We are grateful to Mika Straka, Joffa Applegate, Renaud Lambiotte, Blas Kolic, Michael Osborne, Frank Neffke and the INET Complexity Economics Research Group for their valuable discussions and feedback. R. Maria del Rio-Chanona would also like to acknowledge funding from the Conacyt-Sener doctoral scholarship. This work was supported by Partners for a New Economy, the Oxford Martin School Programme on Technological and Economic Change, the Oxford Martin School Programme on the Post-Carbon Transition and Baillie Gifford.

\newpage







 


\newcommand{\captionabove}[2][]{%
    \vskip-\abovecaptionskip
    \vskip+\belowcaptionskip
    \ifx\@nnil#1\@nnil
        \caption{#2}%
    \else
        \caption[#1]{#2}%
    \fi
    \vskip+\abovecaptionskip
    \vskip-\belowcaptionskip
}

\renewcommand{\theequation}{S\arabic{equation}}
\renewcommand{\thefigure}{S\arabic{figure}}
\renewcommand{\thetable}{S\arabic{table}}
\renewcommand{\thesection}{S\arabic{section}}


\topmargin 0.0cm
\oddsidemargin 0.2cm
\textwidth 16cm 
\textheight 21cm
\footskip 1.0cm

\baselineskip16pt


\maketitle 



\section*{Supplementary Material. Automation and occupational mobility: A data-driven network model }
\subsubsection*{R. Maria del Rio-Chanona, Penny Mealy, Mariano Beguerisse-D\'iaz, Fran\c{c}ois Lafond, and  J. Doyne Farmer}

\section{Calibration}
To calibrate the model we use fine-grained data when possible and aggregate data when this is not possible.  To calibrate the target labor demand when the shock begins we assume that the labor market is initially in steady state, so that the target labor demand in each occupation is equal to the total employment in that occupation.  We thus assume that $d^\dagger_{i,0} = e_{i,0}$, where $e_{i,0}$ is the average employment in 2016.  We assume that the aggregate target labor demand before the shock.  The measured values of the initial target demand for each occupation are given in Table S2 of the Supplementary Material. Throughout the shock we preserve the condition that the aggregate demand $\sum_i d^\dagger_{i,t}$ remains constant in time.

To calibrate the parameters $\delta_u$, $\delta_v$, and $\Delta t$ (the duration of the time step) we simulate an idealized business cycle and adjust these three parameters to find the best match to the empirical U.S. Beveridge curve from December 2007 to December 2018.  To create the artificial business cycle we assume the aggregate target demand $D_t$ follows a sine wave of the form $D_t = D_0 + a \sin( t/2 \pi T)$, where $D_0$ is the initial demand and $T$ is the period of the business cycle. Based on visual inspection, we assume that the empirical curve has traversed about three quarters of a business cycle between December 2007 and December $2018$.   Thus December 2007 is about a quarter of a cycle past the previous peak and December $2018$ is the new peak.  This gives a period of the oscillation $T = 14.6$ years. (The assumptions about phase do not influence the fit, they only explain our reasoning in choosing $T$).  

We assume the model is at its steady state at the beginning of the simulation, with the initial target demand $d^\dagger_0$ of each occupation matching employment in $2016$ (which is the most recent year where we have data for individual occupations).  We then let the target demand $d^\dagger_{i,t}$ of individual occupations move in tandem according to the sine wave, so that each occupation makes a pro-rata change tracking $D_t$, i.e. $d^\dagger_t = d^\dagger_0 + a sin (t/2\pi T)$ and simulate the model.    

We run an exhaustive search over possible values of the amplitude $a$ of the sine function that determines the amplitude of the business cycle and the parameters $\delta_u$, $\delta_v$, and $\Delta t$.  The objective of the search is to minimize the discrepancy between the model and the empirical Beveridge curve.  As the criterion for goodness of fit we compare the intersection of the enclosed areas. The objective function is
\begin{equation}
    \min_{a, \delta_u, \delta_v, \Delta t} \frac{A_m \cap A_e}{ A_m \cup A_e},
\end{equation}
where $A_m$ is the area enclosed by the Beveridge curve of the model, $A_e$ is the area enclosed by the empirical Beveridge curve, $A_m \cap A_e$ is the intersection of their areas and $A_m \cup A_e$ is the union of their areas.  (To define the area of the empirical Beveridge curve we close it by connecting the starting and endpoints).  The optimal parameters are $a=0.065$, $\Delta t = 6.75$ weeks, $\delta_u=0.016$ and $\delta_v=0.012$. The optimal parameters of the model are reasonably stable with respect to the optimal choice $a = 0.065$.  For example, when we increase $a$ by $10\%$, $\Delta t$ remains constant while $\delta_u$ and $\delta_v$ increase roughly by $6\%$, and when we decrease $a$ by $10\%$, $\Delta t$ and $\delta_u$ remain constant while $\delta_v$ increases by less than $5\%$.

We now calibrate the parameter $r$, which is the probability that a worker changing jobs remains within the same occupation.  We are handicapped by the fact that this is not directly recorded, but we can use data on the annual occupational mobility rate, which is the percentage of workers that change occupations within a year, to infer this indirectly. Previous studies estimated that $19\%$ of workers in the USA changed occupations in a year, i.e. that $81\%$ did not change occupations in a year. A more recent study shows that in the Danish economy the annual occupational mobility rate is $20\%$ \cite{groes2014u}. Therefore, we assume that each year $81\%$ of workers remain in their current occupation and use this to estimate $r$, using the following approach.  

In the previous section, we explained that we use the empirical occupational transitions to incorporate the relative preference with which a worker from occupation $i$ applies to a job vacancy in $j$ (for $i \neq j$). Consistent with this approach (and acknowledging the same caveats), here we use the fact that every year $81\%$ of workers remain in their current occupation to calibrate the preference $r$ with which workers chose to apply to job vacancies in their current occupation.

For simplicity, we consider the following abstraction. We assume that the probability that a worker does not change occupation in one time step is time-invariant and constant across occupations. Then, we observe that in the model, only workers who are unemployed change occupation. Thus, the probability that a randomly chosen worker {\it does not} change occupation in one time step is the probability $1-u$ that she is employed plus the probability $u$ that she is unemployed times the probability $r$ that she does not change occupation, that is $ \left((1 - u) + ur\right)$. Then, the probability $x$ that a worker does not change occupations in $y$ time steps is $x = \left((1 - u) + ur\right)^{y}$. Solving for $r$ implies
\begin{equation}
    r = \frac{x^{1/y} + u  - 1}{u}.
\end{equation}
Our model makes roughly $y = 52/6.75 = 7.7$ time steps in one year.  Assuming $x = 0.81$, $\delta_u = 0.016$ and $u = 0.06$ (the average U.S. unemployment rate since the year 2000) gives the estimate $r = 0.55$.

We have no empirical data to calibrate $\gamma$, which is the rate at which the realized demand adjusts towards the target demand.  However, as we demonstrate in the next section, we have the good fortune that the automation results of the model are fairly insensitive to $\gamma$ across a wide range of reasonable parameters.  We choose $\gamma = 10 \delta_u$.

\begin{table}[ht!]
\centering
\caption{Calibrated parameter values}
\begin{tabular}{p{2.051cm}|p{2.1cm}|p{5.5cm}}
\textbf{Parameter}  & \textbf{Value}   & \textbf{Description}   \\
\hline
$\delta_u$            & 0.0160  & Rate at which employed workers are separated due to the spontaneous process.  \\
$\delta_v$            & 0.0120 & Rate at which employed vacancies are opened due to the spontaneous process. \\
$\gamma$            & 0.160  & Rate at employed workers and vacancies are separated or opened due to market adjustment towards the target demand.\\
$\Delta t$            & 6.75 &  Duration of a time step in units of weeks.  \\
$r$                 & 0.55 & Probability that a worker stays in the same occupation  \\
\end{tabular}
\label{tab:parameter_values}
\end{table}

\section{Robustness} \label{sec:SI-robustness}
In this section we test the performance of our approximation and the robustness of our result for different measurements of average unemployment and long-term unemployment, automation shocks, and parameter values. We also study the behaviour of the Beverdige curve under different parameters. This section is structured as follows. 
First, we explore how well our approximations perform with respect to the simulation of the model (this is complementary to the analytical results for our approximations in section \ref{sec:approximations}). Second, we discuss different forms in which we can measure the change in unemployment and long-term unemployment. Third, we test different automation shocks one using the Brynjolfsson et al. estimates and others using the Frey and Osborne shock but losing the hypothesis that the aggregate demand remains constant. Fourth, we explore how different assumptions on the duration of the automation shock and measuring period affect our results. Fifth we explore how different values of $\gamma$ affect our results. 

\subsection{Simulations vs approximation at the occupation level} \label{sec:SI-approximation_checks}
\begin{figure}[h!]
\begin{center}
\includegraphics[width=0.98\textwidth]{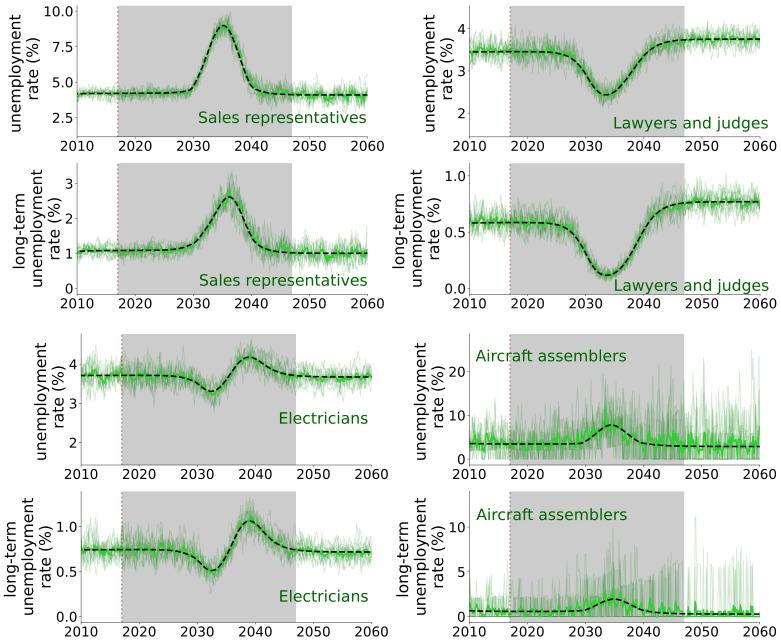}
\caption{\textbf{Unemployment rate at the occupation level simulations and numerical solution} We compare the unemployment and long-term unemployment from the average of the simulations (solid line) and the numerical solution (dashed line). We also show in transparent lines the 10 simulations. Each simulation uses 1.5 Million agents and we average over 10 simulations.}
\label{fig:num_vs_sim_occ}
\end{center}
\end{figure}

We show how our approximations compare with simulations at the occupation. Additionally, we 
discuss the different reaction to the automation shock different occupations have. In particular, we focus on four occupations that we use as examples: sales representatives, lawyers and judges, and electricians. For each, we compare the average of 10 simulations with our numerical solution. As shown in Fig. \ref{fig:num_vs_sim_occ} our approximate solution closely matches the average for all occupations. Because of computational constraints, we run the simulation with 1.5 Million agents which corresponds to one hundredth of the labor force. If we were to run the simulations with the full labor force is ($150$ Million workers) our approximation would improve further.

We focus on four occupations, the sales representatives, lawyers and judges, electricians, and aircraft assembles, whose real employment is $792,000$, $652,000$, $1.1$ million, and $6609$ respectively. We note that aircraft assembles is the second occupation with the smallest employment. Since we run the numerical simulations with a hundredth of the real labor force, in our simulations the target demand for each occupation is $7,920$, $6,520$, $11,000$ thousands, and $66$ respectively. As shown in Fig. \ref{fig:num_vs_sim_occ} our approximations match the average unemployment and long-term unemployment rate of each occupation. Noticeably, the fluctuations are much large for aircraft assemblers, the occupation that has smallest target demand. When we run the simulations with 1.5 million agents all occupations (except Motion picture projectionists) have a target demand above $50$ and most occupations continue a target demand above $50$, therefore we can conclude that our approximations work well for cities with a labor pool above 1.5 million. 
Furthermore, if we ran the simulations with the full labor force, we expect that our approximations would closely match the average behaviour of the unemployment rates for all occupations


We also comment the different behaviour of the occupations. The sales representatives, that are likely to be automated, increase their unemployment rate during the shock. Then, the unemployment rate returns to a steady-state with a similar value to the previous one. Instead, lawyers and judges, who are unlikely to be automated, decrease their unemployment rates during the shock. However, after the shock the steady-state unemployment is higher than it was before the shock. Finally, the electricians who are unlikely to be automated initially decrease their unemployment rate but then increase it during the automation shock. We explain this behaviour as follows. During the first part of the automation shock more electrician vacancies open decreasing unemployment. Nevertheless, the automation shock also causes workers of nearby occupations to become unemployed. As the automation shock continues to separate workers of neighboring occupation, many of these unemployed workers apply for the electrician vacancies causing the electrician's unemployment rate to increase.

\subsection{Measuring the impact of automation}\label{sec:SI-measuring}
In the main text, we define the occupation-specific average unemployment and average long-term unemployment as follows:
\[
u_{i,\text{average}} (T)= \frac{100}{T} \frac{\sum_{t\in T} u_{i,t}}{\sum_{t\in T} (u_{i,t} + e_{i, t})}
\]
and
\[
u_{i,\text{average}}^{(\geq \tau)}(T)= \frac{100}{T}\frac{\sum_{t\in T} u^{(\geq \tau)}_{i,t}}{\sum_{t\in T} (u_{i,t} + e_{i, t})}.
\]
However, there are other measurements of unemployment and long-term unemployment one can compute. For example, we can compute the unemployment rate at each time step of the automation period and then take the average. Thus, we define the alternative unemployment rate and the alternative long-term unemployment rate by
\[
u_{i,\text{alternative}}(T)=  \sum_{t\in T} \frac{ u_{i,t}}{(u_{i,t} + e_{i, t})}
\]
and
\[
u_{i,\text{alternative}}^{(\geq \tau)}(T)=  \sum_{t\in T} \frac{ u^{(\geq \tau)}_{i,t}}{(u_{i,t} + e_{i, t})}.
\]
In Fig. \ref{fig:robustness_gamma_FO} we compare the change in the average unemployment and long-term unemployment rate with the change in the alternative unemployment and long-term unemployment rate. On the top and right we show the change in the average unemployment rate in green and the change in the alternative unemployment rate in cyan. On the bottom and right we do the same for the long-term unemployment rate. Both these plots show that there is a strong overlap between the average change and the alternative change. 

For a better visualization we plot the change in the average unemployment rate vs the alternative unemployment rate on the top left panel. On the bottom left we plot the change in the average long-term unemployment rate vs the alternative long-term unemployment rate. We observe that almost all occupations lie close to the identity line with the exception of occupations that have low employment (small circles) and are highly likely to be automated (red color). The reason behind this discrepancy is that occupations that are highly likely to be automated and have low employment both increase the number of unemployed workers and also decrease the share of employment (due to the structural change). Thus, the ratio between the two, which is considered by the alternative unemployment rates, the increases considerably. Contrarily, when we measure the average unemployment rate, the initial share of employment prevents the sharp increase. However, both measurements exhibit the network effects -- occupations with similar automation probabilities have different percentage change in their unemployment and long-term unemployment rates.

\begin{figure}
\begin{center}
\includegraphics[width=0.9\textwidth]{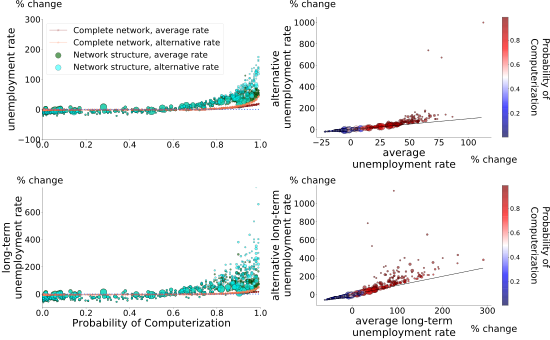}
\caption{\textbf{average and alternative unemployment and long-term unemployment rates} \textbf{Left.} We show the percentage change in the average unemployment and long term unemployment rates in green and the percentage change in the alternative  unemployment and long term unemployment rates in cyan. \textbf{Right.} We plot the percentage change of the average unemployment and long-term unemployment rate vs the percentage change of the alternative unemployment and long-term unemployment rate}
\label{fig:period_av_FO}
\end{center}
\end{figure}

\subsection{Brynjolfsson et al. shock}

Brynjolfsson et al. took a different approach than Frey and Osborne to assess the automatability of occupations. Taking advantage of the 8-digit level O*NET classification of occupations based on work activities \cite{brynjolfsson2017can}. (This has 974 occupations). They asked workers from a crowd sourcing platform to rate what they called the {\it suitability for machine learning} for each work activity.  They then used the breakdown of work activities for each occupation to estimate the suitability for machine learning for each occupation. The suitability for Machine Learning score is based on a five point scale \cite{brynjolfsson2018SML}. We normalize this measure by dividing it by 5, so that it is in a range from zero to one. Most occupations have at least some tasks that are suitable for machine learning, but few, if any, have all tasks suitable for machine learning. This suggests that many jobs will be re-designed rather than destroyed.  

\begin{figure}
\begin{center}
\includegraphics[width=0.5\textwidth]{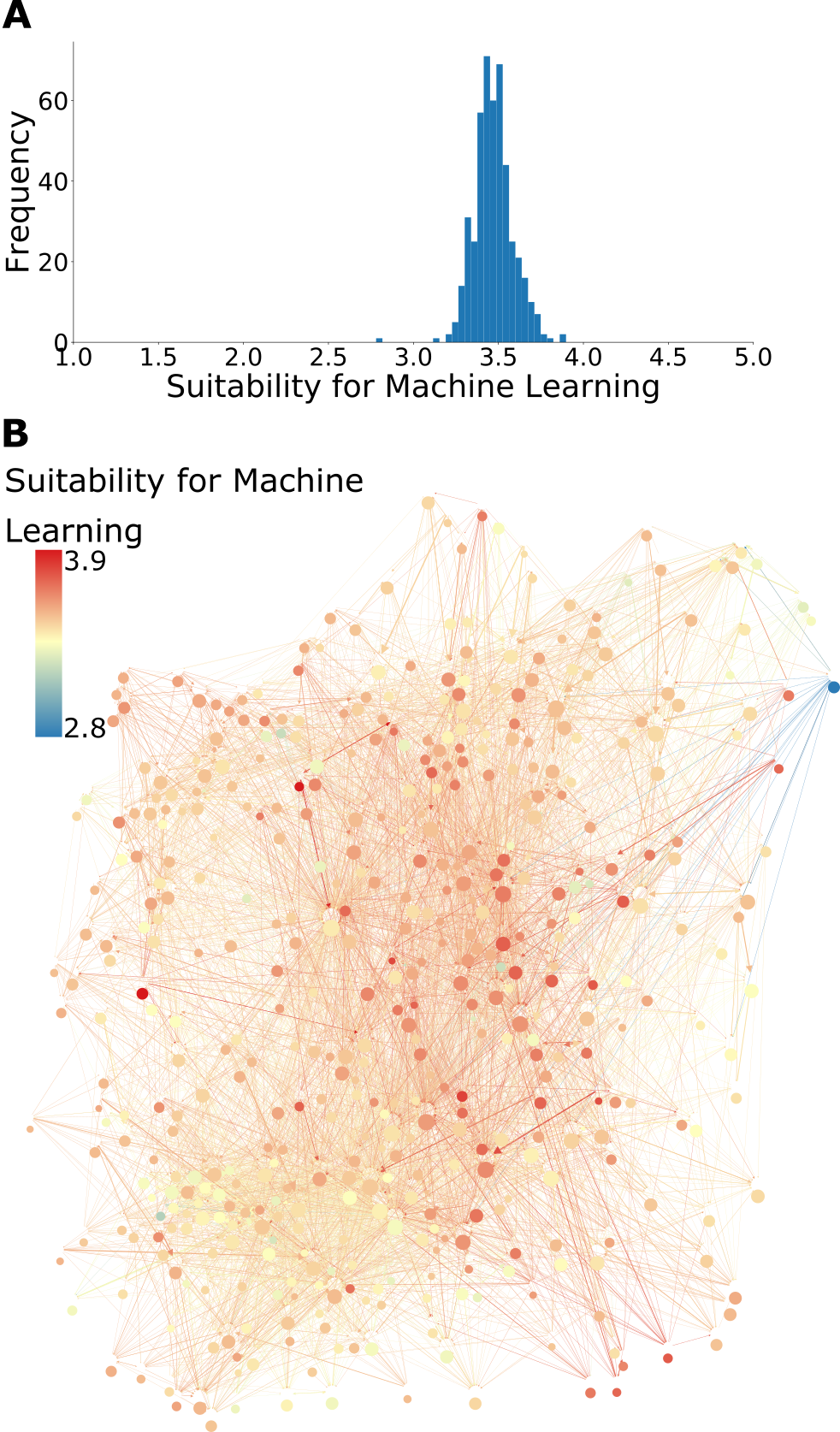} 
\caption{\textbf{Estimates of automatability in the occupational mobility network.} Panel (\textbf{A}) is a histogram of the suitability for machine learning as estimated by Brynjolfsson et al. \cite{brynjolfsson2018SML}. Unlike the Frey and Osborne distributions, the suitability for machine learning distribution is unimodal. Panel (\textbf{B}) shows the occupational mobility network, where nodes represent occupations and links represent possible worker transitions between occupations. The color of the nodes indicates the suitability for machine learning. Red nodes have higher suitability for machine learning and blue nodes have a low one. The size of the nodes indicates the number of employees in each occupation.}
\label{fig:network_sml}
\end{center}
\end{figure}

\begin{figure}
\centering
\includegraphics[width=0.5\textwidth]{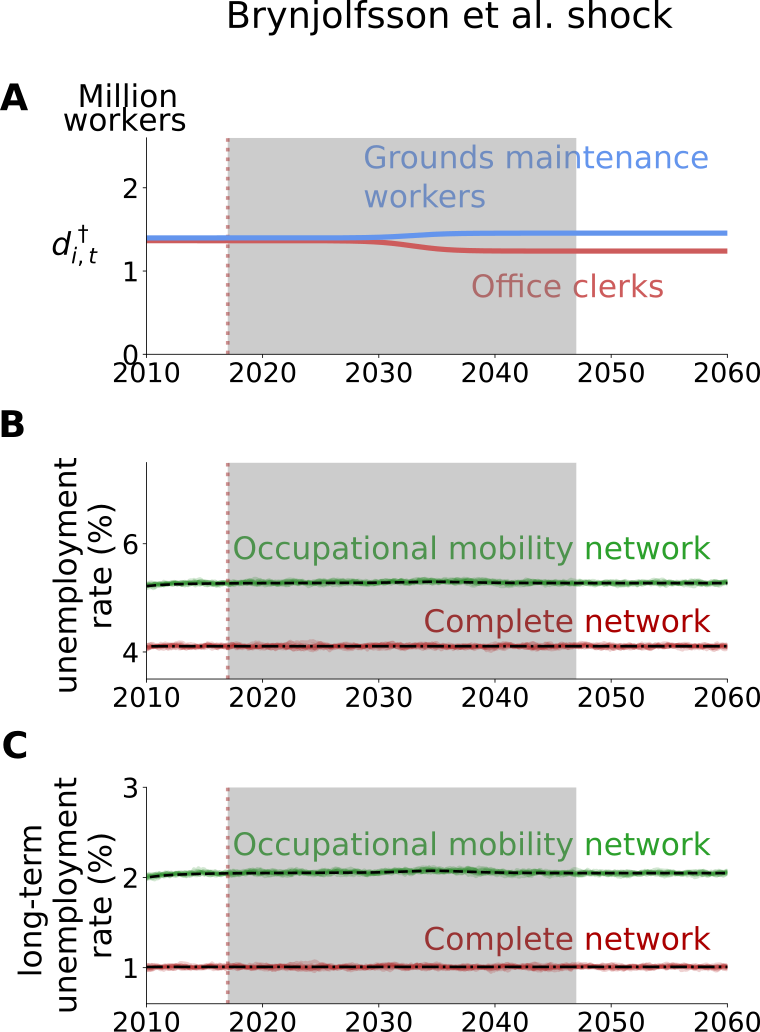}
\caption{\textbf{Aggregate labor market outcomes under the Brynjolfsson et al.  shock.} The grey area denotes the 30 years during which the automation shock takes place. Panel (\textbf{A}) shows the evolution of the target labor demand for two example occupations. The occupation colored in blue has a low suitability for machine learning and the occupation colored in red has a high one. Because the distribution of the suitability for machine learning is more evenly distributed  across occupations that the probability of computerization, the Brynjolfsson et al. shock implies a small change in the target labor demand of most occupations. Panel (\textbf{B}) shows the unemployment rate as a function of time. Dashed lines are our approximations of the expected value (solved numerically) and the solid lines are 10 simulations with 1.5 M agents. Panel (\textbf{C}) shows the long-term unemployment rate as a function of time. As before, dashed lines correspond to the deterministic approximation of Eqs.13--15
and solid lines to the full stochastic model simulation of Eqs.2--4
.}
\label{fig:automation_shock_sml}
\end{figure}

The Brynjolfsson et al. study yielded substantially different results than the Frey and Osborn study. First, these studies differ in their correlation to wages.  
The Frey and Osborne estimates are strongly anti-correlated with wages, whereas the Brynjolfsson et al. estimates have a low correlation with wages. Second, as we see in Fig.1
, the distribution of the Frey and Osborne estimates is wide, whereas the Brynjolfsson et al. distribution has a narrow peak (see Fig. \ref{fig:network_sml}). Since the Frey and Osborne estimates vary substantially between occupations and the Brynjolfsson et al. estimates do not, the corresponding changes in the target labor demand are large for the Frey and Osborne shock but small for the Brynjolfsson et al. shock. (See Fig. 4
B and  \ref{fig:automation_shock_sml}B for examples of how the target labor demand changes for different occupations under the two shocks).

The differences between the Frey and Osborne and the Brynjolfsson et al. shock become clear in the effect they have on employment. The Brynjolfsson et al. shock causes no noticeable change in the aggregate unemployment or long-term unemployment rate (see Fig. \ref{fig:automation_shock_sml}B and C). This is because the Brynjolfsson et al. shock implies small changes in the target demand of occupations (for example, see Fig. \ref{fig:automation_shock_sml}A). 

Although there is no noticeable change in the aggregate unemployment rates, the Brynjolfsson et al. shock still affects occupations disproportionately. This effect depends not only on the suitability for machine learning, but also on the position in the network of each occupation. As we observe in Fig.  \ref{fig:automation_shock_sml}, the change in the long-term unemployment and unemployment is varies substantially for occupations with similar suitability for machine learning. For example, both machinists and avionic technicians have a high $0.70$ suitability for machine learning score, but long-term unemployment for machinists slightly increases, while avionic technicians {\it decrease} their long-term unemployment by more than $20\%$. In other words, our results suggest that retraining efforts would be better spent on machinist than on avionic technicians. 


\subsection{Automation shocks that change the aggregate demand}\label{sec:SI-results_demandchange}
\begin{figure}[H] 
\begin{center}
\includegraphics[width=0.9\textwidth]{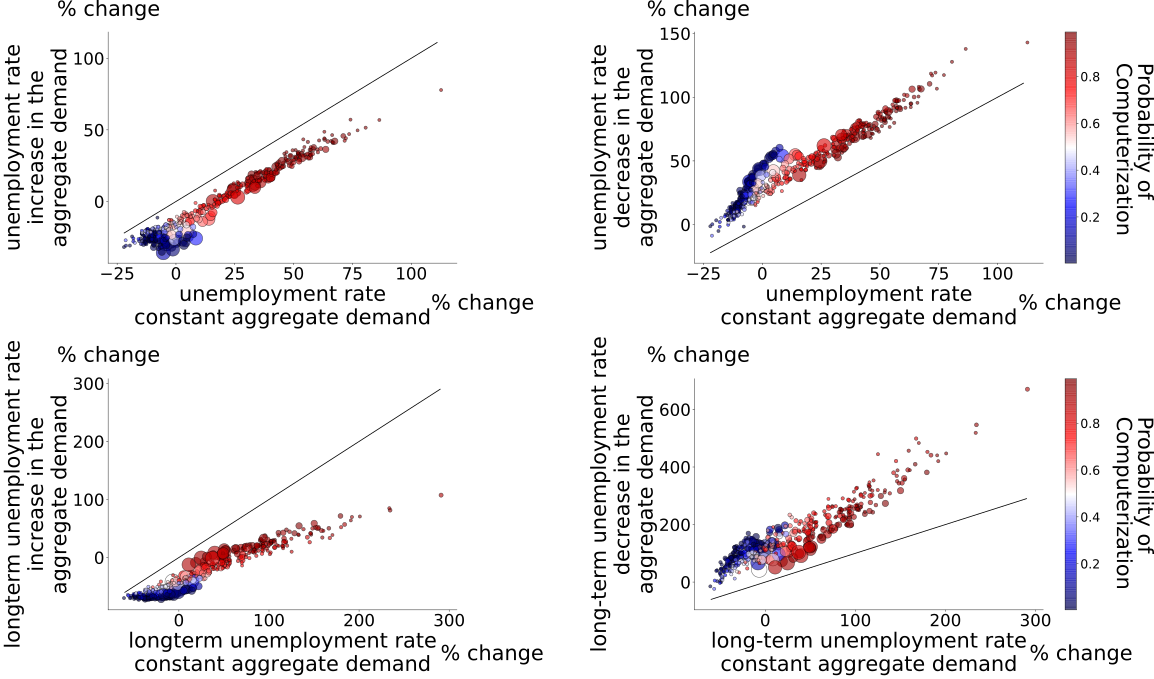}
\caption{\textbf{Frey and Osborne shock with different post-automation target demand scenarios}. In each panel we plot on the x-axis the percentage change in the period unemployment rates when the aggregate demand does not change and on the y-axis the percentage change in the period unemployment rate when the aggregate demand does change. On the left panels we assume the aggregate demand increases by $5\%$ and on the right we assume it decreases by $5\%$.}
\label{fig:different_dstar_FO}
\end{center}
\end{figure}

We assumed that the aggregate demand remains constant after the shock. In this section, we break this assumption and we run our model for an increasing and decreasing aggregate demand of $5\%$. We study the Frey and Osborne automation shock. 
In Fig. \ref{fig:different_dstar_FO} we plot the change in the period unemployment and long-term unemployment rate when the demand changes for each occupations vs the change in the period unemployment  and long-term unemployment rate when the demand remains constant. 
As expected when the aggregate demand increases the percentage change in unemployment and long-term unemployment is lower and the points lie below the identity line. When the aggregate demand decreases, the percentage change in unemployment and long-term unemployment is higher and the points lie above the identity line. While there is a strong correlation between the changes in the unemployment rates when the demand changes and when the demand remains constant, occupations with low automation probabilities (blue dots) lie further away from the identity line. This result means that the structural part of the automation shock mostly affects the occupations with high estimates of automation. When we include a change in the aggregate labor demand then occupations with low automation estimates are also affected considerably. 

\subsection{Automation time and adoption rate}\label{sec:SI-robustness-time}
\begin{figure}
\begin{center}
\includegraphics[width=0.98\textwidth]{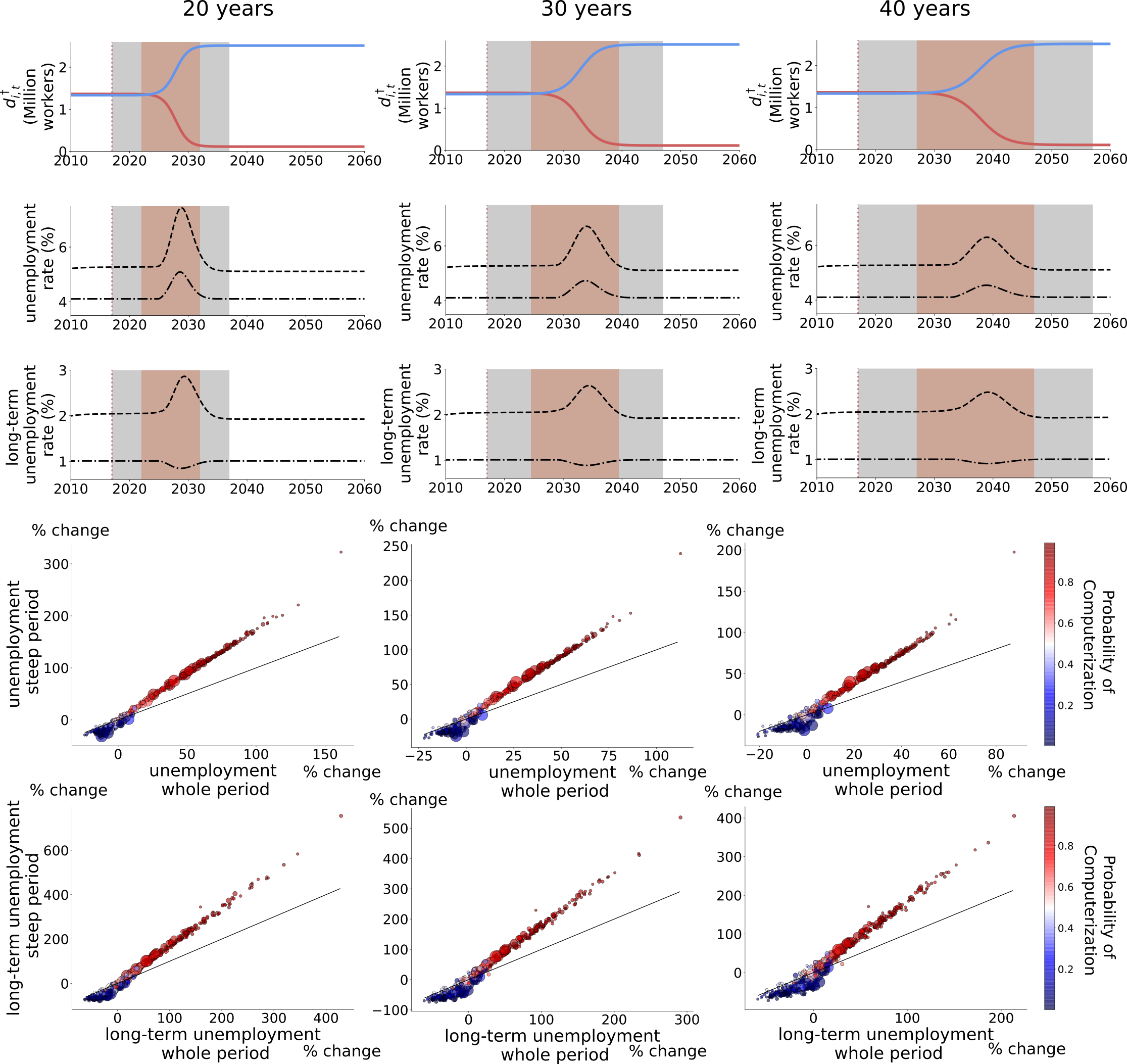}\\
\caption{\textbf{Shock duration and measuring windows effect on the measurements on unemployment rates} \textbf{Top.} For different duration of the automation shock (20, 30 and 40 years) we show target demand of two occupations with similar demand level (Childcare workers and officer clerks), the unemployment, and long-term unemployment rates. The grey area denotes the whole period of automation, meaning that the target demand has reached the automation level within a $1\times10^{-4}$ tolerance. The coral area denotes the sharp transition period which is middle steepest part of the Sigmoid shock. \textbf{Bottom.} For each occupation we plot the percentage change during the whole transition period vs the percentage change during the sharp transition period. Occupations are colored by their automation probability.}
\label{fig:robustness_time}
\end{center}
\end{figure}

In this section we discuss how our results change when we assume a different duration of the automation shock. We assume that automation happens within $20$ and $40$ years, instead of $30$. We measure the change in unemployment during the whole automation period and during the \textit{steep} transition period. We define the steep transition period as the middle part of the automation period when of the sigmoid is steepest. In Fig. \ref{fig:robustness_time} we highlight the whole automation period with a grey area coloring and the steep automation period by a coral shadowing. 

As expected, the shorter the automation period is, the larger the increase in the aggregate unemployment and long-term unemployment rates (see top panels of Fig. \ref{fig:robustness_time}). On the bottom panels of Fig. \ref{fig:robustness_time} we plot the percentage change in the unemployment and long-term unemployment rates of each occupation during the whole automation period vs the percentage change of the unemployment rates of each occupations during the steep automation period. There is a strong correlation between the change in unemployment during the whole transition period and during the steep transition period. However, during the steep automation period the percentage change in the unemployment rate is more extreme than during the whole automation period. Namely, occupations with high automation level have a higher percentage change in the unemployment and long-term unemployment rate during the steep automation period than during the whole automation period. Likewise, occupations with low automation level tend to decrease their unemployment and long-term unemployment rate more during the steep automation period than during the whole automation period


\subsection{Results for different values of $\gamma$} \label{sec:SI-robustness-gamma}


Here, we check the robustness of our results with respect to different values of the $\gamma$ parameter. As explained in the main text, $\gamma$ represents the rate at which the market adjust towards the target demand. In our model $\delta_u$ and $\delta_v$ represent the probability that workers (vacancies) and separated (open) due to random events; and $\gamma$ is a adjustment rate. Therefore,  we expect that $\gamma \geq \delta_u$ and $\gamma \geq \delta_v$. In the main text we use $\gamma = 10\delta_u$ as a reference point. In this section we explore how our results change for different values. In particular we test for $\gamma = 5\delta_u$ and $\gamma = 20\delta_u$. We chose these ranges since there is little change in the results for larger values of $\gamma$ and for lower values, we obtain unreasonably high values of the unemployment rate at the aggregate level (more than $15\%$). 

\begin{figure}
\begin{center}
\caption{Expected change in unemployment and long-term unemployment with different $\gamma$ parameters.}
\includegraphics[width=0.75\textwidth]{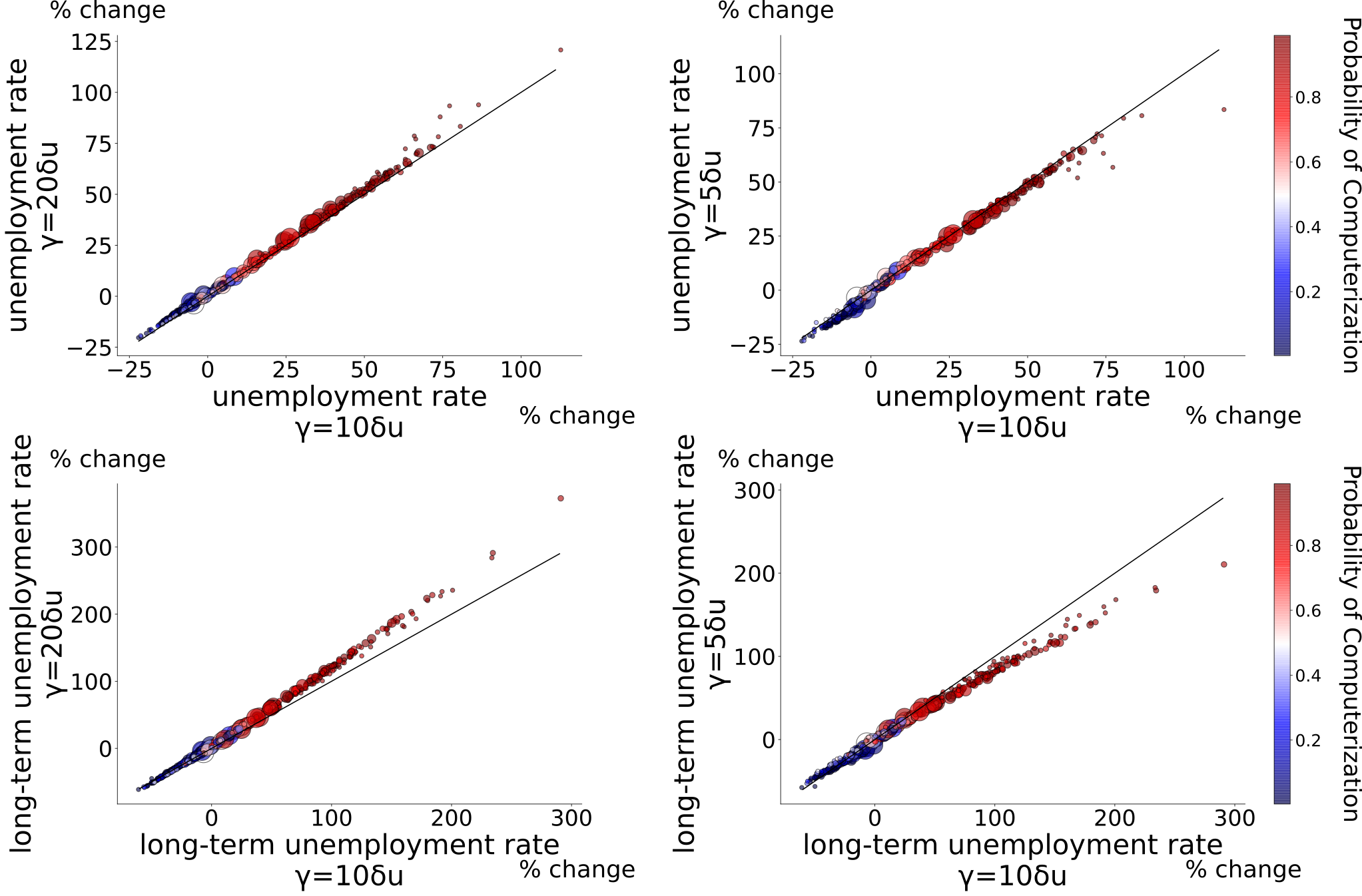}
\caption{\textbf{ Change in unemployment and long-term unemployment with different values of gamma}  Top panels show the change in unemployment rates vs the automation probability. The bottom panels show the change when $\gamma = 5\delta_u$ and  $\gamma = 20\delta_u$ on the y axis and on the x-axis when $\gamma = 10\delta_u$.}
\label{fig:robustness_gamma_FO}
\end{center}
\end{figure}

In Fig. \ref{fig:robustness_gamma_FO} we plot the percentage change in the unemployment rate using $\gamma = 10\delta_u$ (our benchmark) vs the percentage change in unemployment rate when  $\gamma = 5\delta_u$ and  $\gamma = 20\delta_u$ respectively. Our results show that the changes are very similar although as $\gamma$ increases so does the increase in unemployment and long-term unemployment for occupations that are likely to be automated. These results are not surprising, since the larger $\gamma$ is the faster the adjustment towards the target demand and thus sharper the shock.

\onecolumn

\section{The dynamics of the Beveridge curve} \label{sec:SI-robustness}
\begin{figure}
\begin{center}
\includegraphics[width=0.75\textwidth]{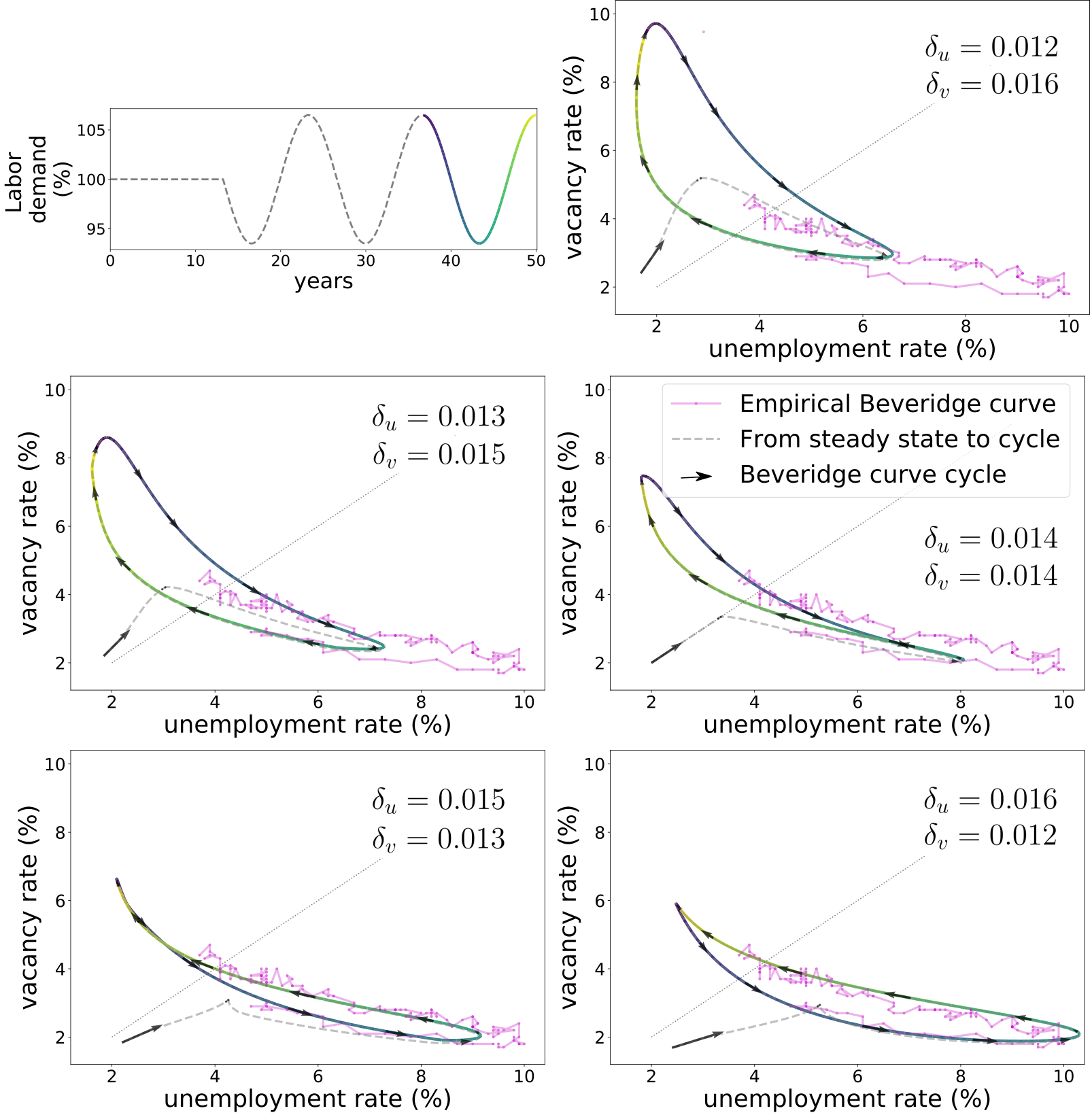}
\caption{\textbf{Beveridge curve dynamics for the occupational mobility network.} On the top left panel we show the aggregate demand. The grey part corresponds to the steady-state and transition to the business cycle. The purple/blue part corresponds to the recession period, while the green/yellow part to the recovery period. The following 5 panels show the dynamics of the model's Beveridge curve under different parameter choices. We observe that when $\delta_u > \delta_v$ the curve cycles counter-clockwise, while when  $\delta_u \leq \delta_v$ the curve cycles clockwise.  We also show the empirical Beveridge curve in magenta for reference.}
\label{fig:Bevcurve_omn}
\end{center}
\end{figure}

\begin{figure}
\begin{center}
\includegraphics[width=0.75\textwidth]{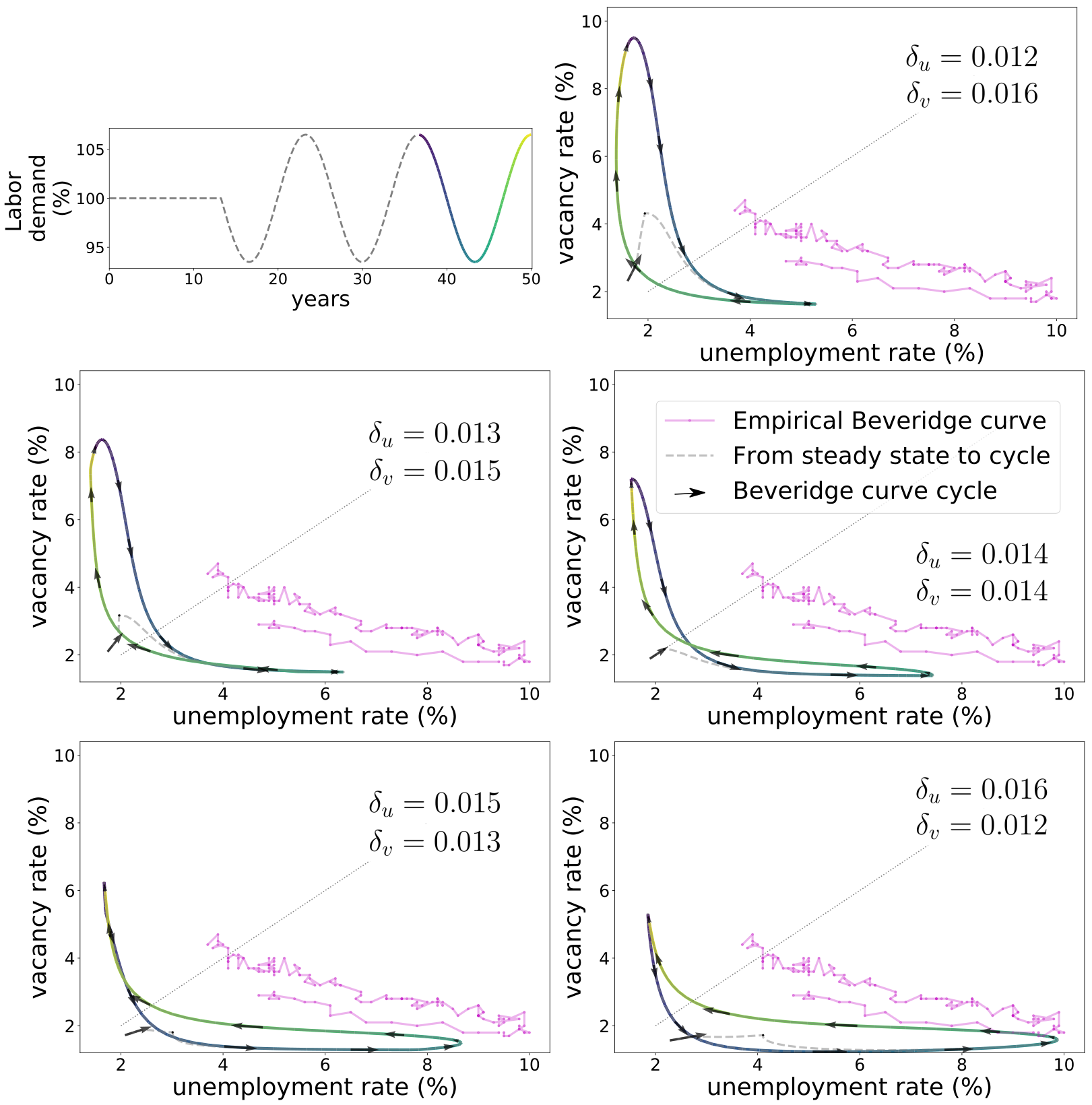}
\caption{\textbf{Beveridge curve dynamics for the complete network} On the top left panel we show the aggregate demand. The grey part corresponds to the steady-state and transition to the business cycle. The purple/blue part corresponds to the recession period, while the green/yellow part to the recovery period. The following 5 panels show the dynamics of the model's Beveridge curve under different parameter choices. We observe that when $\delta_u \geq \delta_v$ the curve cycles counter-clockwise, while when  $\delta_u < \delta_v$ the curve cycles clockwise.  We also show the empirical Beveridge curve in magenta for reference.}
\label{fig:Bevcurve_kn}
\end{center}
\end{figure}

We explore the dynamics of the Beveridge curve by varying $\delta_u$ and $\delta_v$. We keep all other parameters fixed to the values used to fit the Beveridge curve ($a = 0.065$, $\Delta t = 6.75$ weeks, and $\gamma_u$ = $\gamma_v$ = $\gamma = 0.16$). As before, we use a $sine$ wave to model business cycle dynamics. In the top left of Figs.\ref{fig:Bevcurve_omn} and  Fig.\ref{fig:Bevcurve_kn}) we show the dynamics of the aggregate target demand. We start with a constant target demand and then introduce the business cycle dynamics. We show the first part of the dynamics with a dashed line to mark the transition between a constant target demand and an oscillating target demand. Then, we plot in color-scale the dynamics of a business cycle, the purple/blue part corresponds to the recession period, while the green/yellow part to the recovery period. 

We test five different parameter options for $\delta_u$ and $\delta_v$. Starting from the calibrated values $\delta_u = 0.016$ and $\delta_v = 0.012$, we gradually decrease $\delta_u$ by $0.001$ and increase $\delta_v$ by $0.001$ until $\delta_u = 0.012$ and $\delta_v = 0.016$. This yields 5 different cases, which we show in Fig.\ref{fig:Bevcurve_omn} for the occupational mobility network and in Fig.\ref{fig:Bevcurve_kn} for the complete network.

We find that the Beveridge curve first reduces its enclosed area, then it changes its cycling direction from counter-clockwise to clockwise, and finally, it increases its enclosed area. For the two networks and for the five cases we study, we observe that when $\delta_u > \delta_v$ the curve cycles counter-clockwise. On the contrary, when $\delta_v > \delta_u$ the curve cycles clockwise. However, when $\delta_u = \delta_v$ the network determines the the direction of the cycle -- the occupational mobility network shows clockwise cycles while the complete network shows an "8"-shaped curve, where the bottom part cycles counter-clockwise and the upper part clockwise. These results suggest that for some similar values of $\delta_u$ and $\delta_v$, which depend on the network structure, the curve flips and starts to exhibit the opposite cycling direction. We also observe that the network structure affects the area enclosed by and the position of the curve (see differences between  Fig.\ref{fig:Bevcurve_omn} and  Fig.\ref{fig:Bevcurve_kn}).  



\section{Mathematical derivations and approximations}

\subsection{Deterministic approximation for large population}\label{sec:approximations}
To obtain Eqs. 13 -- 15 we must first compute the expected value of separated workers and opened vacancies, which is straightforward, and then compute the expected flow of workers, which requires more steps. 

\subsubsection{Separated workers and opened vacancies.}
The number of separated workers and opened vacancies follow a binomial distributions, meaning that their expected value is just the success rate times the number of trials.  Thus, from Eqs. 5, 9 and 11 
we can compute the expected number of separations conditioned on the unemployment, and similarly, from Eqs. 6, 10 and 12 
we can compute the expected number of vacancies.  This gives
\begin{eqnarray}
 \label{eq:expected_omega}
   \bar{\omega}_{i,t+1}  &=\pi_{u,i,t}\bar{e}_{i,t}  =  &\delta_u  \bar{e}_{i,t} +  (1 - \delta_u) \gamma \max \big\{0,  \bar{d}_{i,t} - d_{i,t}^\dagger \big\}, \\
 \bar{\nu}_{i,t+1} & = \pi_{v,i,t}\bar{e}_{i,t}  =  &\delta_v  \bar{e}_{i,t} +  (1 - \delta_v)\gamma \max \big\{0, d_{i,t}^\dagger - \bar{d}_{i,t}  \big\}.
\end{eqnarray}
\subsubsection{Flow of workers}
The labor flow $f_{ij,t+1}$ is equal to the number of workers from occupation $i$ applying to occupation $j$, $s_{ij,t+1}$, multiplied by the probability $p_{j,t+1}$ that each application is successful. (All applications are accepted with uniform probability, so $p$ does not depend on $i$). The expected value is
\begin{equation}
 \bar{f}_{ij,t+1} = E\left[ s_{ij,t+1} p_{j,t+1} | \mathbf{u}_{i,t}, \mathbf{v}_{i,t}, \mathbf{e}_{i,t} \right].
 \label{eq:f_ij}
\end{equation}
Letting the total number of applications $s_{j,t+1}$ to occupation $j$ be
\begin{equation}
    s_{j,t+1} = \sum_k s_{kj,t+1},
\end{equation}
the fraction $p_{j,t+1}$ of successful applications is the ratio of the number of vacancies $m_{j,t+1}$ that successfully match to the total number of applications, i.e. 
\begin{equation}
    p_{j,t+1} = m_{j,t+1}/s_{j,t+1}.
    \label{eq:p_j}
\end{equation}

\paragraph{Matching.}
To simplify upcoming calculations that require derivatives, it is convenient to express the number of matches $m_{j,t+1}$ in terms of the exponential function. This is a standard approximation, and the derivation we present is based on \cite{stevens2007new}. Recall that $m_{j,t+1}$ is the number of vacancies that successfully match with a job applicant. Since employees hire a worker uniformly at random from the pool of applicants, then  $m_{j,t+1}$ is equal to the number of job applications that receive at least one job application. 

An unemployed worker who applies for a job in occupation $j$ with $v_{j, t}$ vacancies, will apply to a particular vacancy with probability $\frac{1}{v_{j, t}}$.
 Thus the probability that the worker does {\it not} send her application to that vacancy is $1 - \frac{1}{v_{j, t}}$.  For $s_{j, t}$ unemployed workers sending applications to occupation $j$, the probability that a particular vacancy does not receive an application is $(1 - \frac{1}{v_{j, t}})^{s_{j, t}}$. Since each vacancy receiving an application hires one worker, the expected number $\bar{m}_{j,t+1}$ of successful job applications is 
\begin{equation}
  \bar{m}_{j, t+1} = v_{j, t}\left(1 - (1 - \frac{1}{v_{j, t}})^{s_{j, t}}\right).\nonumber
\end{equation}
Using the approximation that $(1 - x)^y \approx  e^{-xy}$  for large $x$ and $y$ we obtain
\begin{equation}
   \bar{m}_{j, t+1} = v_{j,t}(1 - e^{- s_{j,t+1}/v_{j,t}}).
  \label{eq:m_approx}
\end{equation}

\paragraph{Taylor approximation for flow of workers}
From Eqs. \ref{eq:f_ij} and \ref{eq:p_j} we observe that $\bar{f}_{ij,t+1}$ depends on the variables $s_{ij,t+1}$, $m_{j,t+1}$ and $s_{j,t+1}$. These variables are not independent, but here we show that in the large $L$ limit we can approximate $\bar{f}_{ij,t+1}$ as
\begin{equation}
 \bar{f}_{ij,t+1} \approx  \bar{s}_{ij, t+1} \bar{v}_{j, t} \frac{(1 - e^{-\bar{s}_{j, t+1} /\bar{v}_{j, t}})}{\bar{s}_{j, t+1}}.
\end{equation}
To compute $\bar{f}_{ij,t+1}$ it is useful to define
\begin{equation}
 s_{j\backslash i, t+1} \equiv \sum_{k \neq i} s_{kj, t+ 1},
 \label{eq:s_backslash}
\end{equation}
which is the number of applications occupation $j$ receives from all unemployed workers except those from occupation $i$. Note that $s_{j,t+1} =  s_{j\backslash i, t+1} + s_{ij,t+1}$. Using this fact, and Eq. (\ref{eq:f_ij}) we define the following multivariate form of the flow of workers
\begin{equation}
     \bar{f}_{ij,t+1} = g(s_{ij, t+ 1}, s_{j\backslash i, t+1}) \equiv s_{ij, t+1} v_{j, t} \frac{(1 - e^{- (s_{ij, t+1} +  s_{j\backslash i, t+1})/v_{j, t}})}{s_{ij, t+1} +  s_{j\backslash i, t+1}}.
    \label{eq:h}
\end{equation}
This definition will allow us to do a multivariate Taylor expansion of the function $g$ around the expected value of $s_{ij, t+ 1}$ and $s_{j\backslash i, t+1})$.

Recall that, for fixed $i$, the random variables $s_{ij,t+1}$ follow a multinomial distribution with $u_{i,t}$ trials and probabilities $q_{ij, t+1}$ for $j = 1,...,n$. This means that $s_{ij, t+1}$ and $s_{il, t+1}$ are drawn from the same realization of the multinomial distribution, and are therefore correlated. However, $s_{ij, t+1}$ and $s_{kj, t+1}$ are drawn from different realizations (and distributions); thus they are independent. 

In other words, the number of workers from occupation $i$ that apply to occupation $j$ is correlated with the number of workers from occupation $i$ that apply to occupation $l$ -- if all $u_{i, t}$ workers apply to occupation $j$ it means that no workers from $i$ applied to occupation $l$. However, since workers do not coordinate when sending applications, the fact that many or few workers from occupation $i$ apply to occupation $j$ says noting about the number of workers from occupation $k$ that applied to occupation $j$. Of course, this is conditional on $v_{j,t}$, the number of vacancies in occupation $j$ at the previous time step.

It follows from the fact that $s_{ij, t+1}$ and $s_{kj, t+1}$ are independent and from Eq. (\ref{eq:s_backslash}), that $s_{ij, t+1}$ is independent from $s_{j\backslash i, t+1}$. Furthermore, in the limit of a large number of agents, $u_{i,t}$ is large, and the standard deviation of $s_{ij,t+1}$ is small in comparison to the average. The same is true for $s_{j\backslash i,t+1}$; we can therefore expand $g(s_{ij, t+ 1}, s_{j\backslash i, t+1})$ in a Taylor series around the expected value of $s_{j,t+1}$ and $s_{j\backslash i, t+1}$ as follows,
\begin{eqnarray}
\bar{g}(s_{ij, t+ 1}, s_{j\backslash i, t+1}) & = & g(\bar{s}_{ij, t+ 1},  \bar{s}_{j\backslash i, t+1}) \nonumber \\
 &+ & \frac{1}{2} \frac{\partial ^2}{ \partial s_{ij, t+ 1}^2}\left( g(\bar{s}_{ij, t+ 1},  \bar{s}_{j\backslash i, t+1})\right) \text{Var}[s_{ij, t+ 1}] \nonumber \\  
 &+ &  \frac{1}{2} \frac{\partial ^2}{ \partial s_{j\backslash i, t+1}^2} \left(h(\bar{s}_{ij, t+ 1}, \bar{s}_{j\backslash i, t+1})\right) \text{Var}[s_{j\backslash i, t+1}] + \dots
  \label{eq:taylor}
\end{eqnarray}
Next, we now show that, in the limit of a large number of agents, the second and third terms are negligible in comparison to the first term. Exclusively for this derivation, we introduce the notation $v \equiv v_{ij,t} $,  $x \equiv s_{ij, t+ 1} $ and $y \equiv s_{j\backslash i, t+1}$. Furthermore, we denote the expected value of a variable $x$ by $\mu_x$ and the variance by $\sigma_x^2$. Using this notation and taking the partial derivatives from Eq. (\ref{eq:h}), we obtain
\begin{eqnarray}
\bar{h}(x,y) & = & h(\mu_x, \mu_y) + \sigma_x^2 \left[\left( \frac{v\mu_x}{(\mu_y + \mu_x)^3} - \frac{v}{(\mu_y + \mu_x)^2}\right)\left( 1 - e^{-(\mu_x + \mu_y)/v} \right)   \right. \nonumber \\
 &+ &  \left. \left( \frac{1}{\mu_y + \mu_x} - \frac{\mu_x}{(\mu_y + \mu_x)^2} \right)e^{-(\mu_x + \mu_y)/v} - \frac{1}{2}\frac{\mu_x}{v(\mu_y + \mu_x)}e^{-(\mu_x + \mu_y)/v}  \right] \nonumber \\  
 &+ & \sigma_y^2 \left[ \frac{v\mu_x(1 - e^{-(\mu_x + \mu_y)/v})}{(\mu_x + \mu_y)^3} -  \frac{\mu_x}{(\mu_x + \mu_y)^2}e^{-(\mu_x + \mu_y)/v}\right.  \nonumber \\
 &- & \left. \frac{1}{2} \frac{\mu_x}{v (\mu_y + \mu_x)}e^{-(\mu_x + \mu_y)/v} \right],
  \label{eq:taylor_new_notation}
\end{eqnarray}
where we have neglected second order terms in the expansion.

Since $\mu_x$, $\mu_y$, $\sigma_x^2$, $\sigma_y^2$ and $v$ scale linearly with the number of agents $L$, in the limit of a large number of agents these five variables are of the same order of magnitude. It follows from this observation and from Eqs. (\ref{eq:h}) and (\ref{eq:taylor_new_notation}) that, in the limit of large number of agents, the first term of Eq. (\ref{eq:taylor_new_notation}) scales with $L$, while the other terms are of the order of a constant $c$. 
In other words, in the limit of a large number of agents, we can approximate 
\begin{equation}
  \bar{f}_{ij,t+1} = \bar{s}_{ij, t+1} \frac{ \bar{v}^2_{j, t} (1 - e^{-\bar{s}_{j,t+1}/\bar{v}_{j,t}})}{\bar{s}_{j,t+1}},
\label{eq:fij_approx}
\end{equation}
where we have recovered our original notation. The relative error of this approximation is inversely proportional to the number of agents i.e.,  the relative error of our approximation is
$$\left| \frac{ E[f_{ij,t+1}| \mathbf{u}_t, \mathbf{v}_t ; A] - \bar{f}_{ij,t+1}}{E[f_{ij,t+1}| \mathbf{u}_t, \mathbf{v}_t ; A] }\right| = \frac{c}{L + c},$$ 
Since $c$ is a constant, when $L \to \infty$, the relative error tends to zero. 
We further check the quality of our approximations in section \ref{sec:SI-approximation_checks}, where we compare the simulated unemployment and long-term unemployment rate with our analytic solutions at the occupation level.

Substituting Eqs.7 and 8
into Eq.~(\ref{eq:fij_approx}), we can write $\bar{f}_{ij,t+1}$ in terms of the adjacency matrix and the expected values of the state variables as  
\begin{equation}
\bar{f}_{ij,t+1} =  \frac{\bar{u}_{i,t} \bar{v}^2_{j, t} A_{ij} (1 - e^{-\bar{s}_{j,t+1}/\bar{v}_{j,t}})}{\bar{s}_{j,t+1}\sum_k \bar{v}_{k,t} A_{ik}} , 
\label{eq:f_full}
\end{equation}
where
\begin{equation}
\bar{s}_{j,t+1} = \sum_i \frac{\bar{u}_{i,t} \bar{v}_{j, t} A_{ij}}{\sum_k \bar{v}_{k,t} A_{ik}}.
\end{equation}

\subsection{Long-term unemployment}
We can compute the number of long-term unemployed workers in each occupation using Eqs. 13 -- 15 as follows.  The expected number of workers with an unemployment spell of $k$ steps for occupation $i$ at time $t$ is the expected number of workers with an unemployment spell of $k-1$ steps at the previous time step times the probability that a worker of occupation $i$ is not hired. Thus the expected number of unemployed workers of occupation $i$ with an unemployment spell of $k$ time steps $u_{i, t+ 1}^{(k)} $ is given by the recursive equation
\begin{equation}
 \bar{u}_{i, t+ 1}^{(k)} = \bar{u}_{i, t}^{(k-1)} \bigg(1 - \frac{\sum_j\bar{f}_{ji, t}}{\bar{u}_{i, t}}\bigg),
 \label{eq:uk_expectedval_approx}
\end{equation}
with $\bar{u}_{i, 1}^{(1)} = \bar{\omega}_{i,1} = \bar{e}_{i,0} \pi_{u,i,t}$.

The U.S. Bureau of Labor Statistics defines long-term unemployed workers as those who have an unemployment spell of 27 or more weeks. Similarly, in our model the long-term unemployed workers are those who have been unemployed for $\tau$ or more time steps.  
Using Eq. (\ref{eq:uk_expectedval_approx}), we compute the expected number of long-term unemployed workers ($\bar{u}_{i, t+ 1}^{( \geq \tau )}$) by summing over all workers with an unemployment spell of $\tau$ or more time steps
\begin{equation}
 \bar{u}_{i, t+ 1}^{( \geq \tau )} = \sum_{k=\tau}^\infty \bar{u}_{i, t+ 1}^{(k)}.
 \label{eq:ult_expectedval_approx}
\end{equation}


\subsection{Steady state}
In this section, we derive the dynamic equation for the expected value of the realized demand $\bar{d}_{i,t}$ and we discuss that, when the target demand is constant, the master equation of $\bar{d}_{i,t}$ has a fixed point solution. Once $\bar{d}_{i,t}$ reaches the fixed point value, $d_{i,t}$ fluctuates around this value, and we say that the system is at the steady-state (see section \ref{sec:dyn_real_demand}). At the end of this section we use the steady-state solution of $\bar{d}_{i,t}$ to show that, under constant target labor demand, $\bar{e}_{i,t}$, $\bar{u}_{i,t}$ and $\bar{v}_{i,t}$ also have a steady-state solution. Throughout this section we assume the case $\delta_u > \delta_v$; the other case can be solved analogously. 

The realized labor demand is, by definition, the sum of employment and vacancies. Thus, from Eqs. 13--15
$$\bar{d}_{i,t + 1}  = \bar{d}_{i,t} +  (\delta_v - \delta_u)e_{i,t}  + \begin{cases} 
    \gamma_u(1 - \delta_u)(d^\dagger_{i,t} - \bar{d}_{i,t}) & \text{if } \bar{d}_{i,t} \geq d^\dagger_{i,t}  \\
   \gamma_v(1 - \delta_v)(d^\dagger_{i,t} -\bar{d}_{i,t})  & \text{if } \bar{d}_{i,t}< d^\dagger_{i,t}.
  \end{cases} $$
We simplify this expression by defining $\gamma_u' = \gamma_u(1 - \delta_u)$ and $\gamma_v' = \gamma_v(1 - \delta_v)$ as follows,
\begin{equation}
  \bar{d}_{i,t + 1}  = \bar{d}_{i,t}  + (\delta_v - \delta_u)\bar{e}_{i,t}  + \gamma_u'(d^\dagger_{i,t} -  \bar{d}_{i,t}) + (\gamma_v' - \gamma_u') \max \big\{0, d_{i,t}^\dagger -  \bar{d}_{i,t} \big\}.
\end{equation}
Note that when $\delta_u=\delta_v$ and $\gamma_u=\gamma_v=1$ terms cancel out and we obtain $ \bar{d}_{i,t + 1} = d^\dagger_{i,t}$ which corresponds to immediate adjustment.
To find the steady-state we use that the  target labor demand $d_i^\dagger$ is constant and look for the value of the realized demand that satisfies $\bar{d}_{i,t +1} = \bar{d}_{i,t}$. This gives,
\begin{equation}
    (\delta_u - \delta_v)e_{i,t} = \gamma_u'(d_i^\dagger - d_i^*) + (\gamma_u' - \gamma_v') \max \big\{0, d_{i}^\dagger -  \bar{d}_{i}^* \big\}.
\end{equation}
where we have assumed that the number of employed workers has reached a steady-state and has expected value $\bar{e}_i^*$. To determine the term within the maximum function, we must determine if $d_i^\dagger < d_i^*$ or $d_i^\dagger > d_i^*$. If we assume $d_i^\dagger < d_i^*$, we obtain that
$$ d_i^\dagger - d_* =   \frac{\delta_u - \delta_v}{\gamma_u'}e_{i,t}.$$
This is a contradiction since the right hand side is positive, but the left hand side negative (since $\delta_u > \delta_v$).  In the case when $d_i^\dagger > d_i^*$, we find the correct steady-state solution 
$$ d_{i}^* = d_i^\dagger - \frac{\delta_u - \delta_v}{\gamma_v'}e_{i,t}.$$

Doing an analogous analysis for the case $\delta_u < \delta_v$ we obtain the following solution for the steady-state of the realized demand,
\begin{equation}
\bar{d}_{i}^* = 
 \begin{cases} 
     d_{i}^\dagger - \frac{\delta_u - \delta_v}{\gamma_v'} \bar{e}_{i}^* & \text{if } \delta_u \geq \delta_v  \\
   d_{i}^\dagger - \frac{\delta_u - \delta_v}{\gamma_u'} \bar{e}_{i}^* & \text{if } \delta_u < \delta_v.
  \end{cases}
  \label{eq:d_*}
\end{equation}
In other words, when $\delta_u > \delta_v$ the realized demand is lower than the target demand. This happens because when $\delta_u > \delta_v$ (i.e., the probability of separation is higher than the probability of opening a vacancy at random) the adjustment towards the target demand does not fully compensate for asymmetry between the opening and separation rates; thus the steady-state value of the realized demand is lower than the target demand. Similarly, when $\delta_u < \delta_v$, the steady-state value of the realized demand is higher than the target demand. In both cases, the difference between the realized and the target demand at the steady-state is proportional to $|\delta_u - \delta_v|$ and inversely proportional to the adjustment rate $\gamma$.

Next, we show that under constant target labor demand $\bar{e}_{i,t}$, $\bar{u}_{i,t}$ and $\bar{v}_{i,t}$ also have a steady-state solution. Again, we solve the case $\delta_u \geq \delta_v$; the other case can be solved analogously. 
We assume that the realized demand has reached its steady-state value $d_{i}^*$. Then, since the realized demand is the sum of the employment and vacancies, and using Eq. (\ref{eq:d_*}), we obtain the steady-state equation for the number of vacancies,
 \begin{equation}
\bar{v}_i^* = d_{i}^\dagger - \left(1 - \frac{\delta_u - \delta_v}{\gamma_v'}\right) \bar{e}_{i}^*.
  \label{eq:v_*_si}
\end{equation}
From the employment equation (Eq.13
), it follows that 
\begin{equation}
 \bar{e}_{i}^* = \frac{1}{\delta_u} \sum_j \bar{f}_{ji}^*,
 \label{eq:e_*_si}
\end{equation}
where we have used the fact that $\delta_u > \delta_v$ and the steady-state value of the realized demand from Eq. (\ref{eq:d_*}). We can also obtain Eq. (\ref{eq:e_*_si})  
from the vacancy equation Eq. 15

Finally, from the unemployment master equation Eq. 14, 
for a steady-state to exist the unemployment of each occupation, encoded in the vector $\mathbf{u}^*$, must satisfy the following equation,
\begin{equation}
 \frac{1}{\delta_v}\sum_j \bar{f}_{ji}^* =  \frac{1}{\delta_u}\sum_j \bar{f}_{ij}^*.
 \label{eq:u_*}
\end{equation}
In other words, at the steady-state the total inflow of workers into an occupation equals the total outflow of workers of that same occupation.

\subsubsection{Long-term unemployment at the steady-state}

We note that Eq.  (\ref{eq:ult_expectedval_approx}) gives the number of long-term unemployed workers for time $t$. A special case is that of the steady-state when the unemployment rate of each occupation is $u_i^*$. Then, the approximate expected number of unemployed workers with a job spell of $k$ time steps is 
$$
\bar{u}_{i}^{*(k)} = \delta \bar{e}_{i}^* \bigg(1 - \frac{\sum_j \bar{f}_{ij}(\mathbf{\bar{u}}^*, \mathbf{\bar{v}}^*; A)}{\bar{u}_{i}^*}\bigg)^k.
$$
and decays exponentially with $k$.

\section{Shift in the steady state unemployment post-automation}

In this section, we give arguments that support our conjecture that the Frey and Osborne shock causes such persistent effects due to the fact that automation levels of neighboring occupations tend to be similar.

To test our conjecture, we create a surrogate Frey and Osborne shock that randomizes the automation levels of occupations. We do this by randomly shuffling the automation level of each of the 464 occupations, i.e. randomly reassigning each automation level to a new occupation (without replacement).  This preserves the distribution of automation levels but removes any correlation between neighboring occupations.  When we do this the aggregate unemployment rate does not decrease, while the long-term unemployment tends to increase slightly. Thus the most persistent effects disappear when the correlation inherent in the network structure is removed.

To demonstrate that the network correlation structure is indeed the cause, we create another surrogate shock where we randomize relative to the Frey and Osborne shock while intentionally creating a correlation between neighbors. Since occupations of the same O*NET classification typically have high connectivity, we redistribute the probabilities of computerization so that occupations with similar classifications have a similar automation level. We do this by ordering the probabilities of computerization in ascending order and ordering the occupations in ascending order with respect to their occupation code. We then match these to create a surrogate shock with the desired property.  When we impose this shock the post-automation aggregate unemployment and long-term unemployment rate both decrease, confirming our conjecture.

In the main text we show that the steady-state unemployment rate after the Probability of Computerization shock is lower than the steady-state unemployment rate before the shock. We conjecture that this behaviour is most likely caused by the assortativity of the Probabilities of Computerization. That is, in general, the Probability of Computerization of an occupation is similar to the Probability of Computerization of its neighbors in the network. When the system converges to the new steady the majority of the workers are in the occupations that have low Probability of Computerization. Therefore, workers are in occupations that are close by in the network reducing labor market frictions.

To test this hypothesis we use reshuffled versions of the Probability of Computerization shock. That is, we keep the same distribution of the Probabilities of Computerization but assign them to different occupations. We reshuffle the probabilities in two ways: a simple randomized shuffling and an assortative shuffling. For the latter, we redistribute the probabilities of computerization so that occupations of the same classification have a similar automation level. In particular, we sort the probabilities of computerization and match them with the occupations sorted by their classification. This ordering is meaningful since the classification system is designed to have similar occupations close. 

\begin{figure}[H] 
\begin{center}
\includegraphics[width=0.45\textwidth]{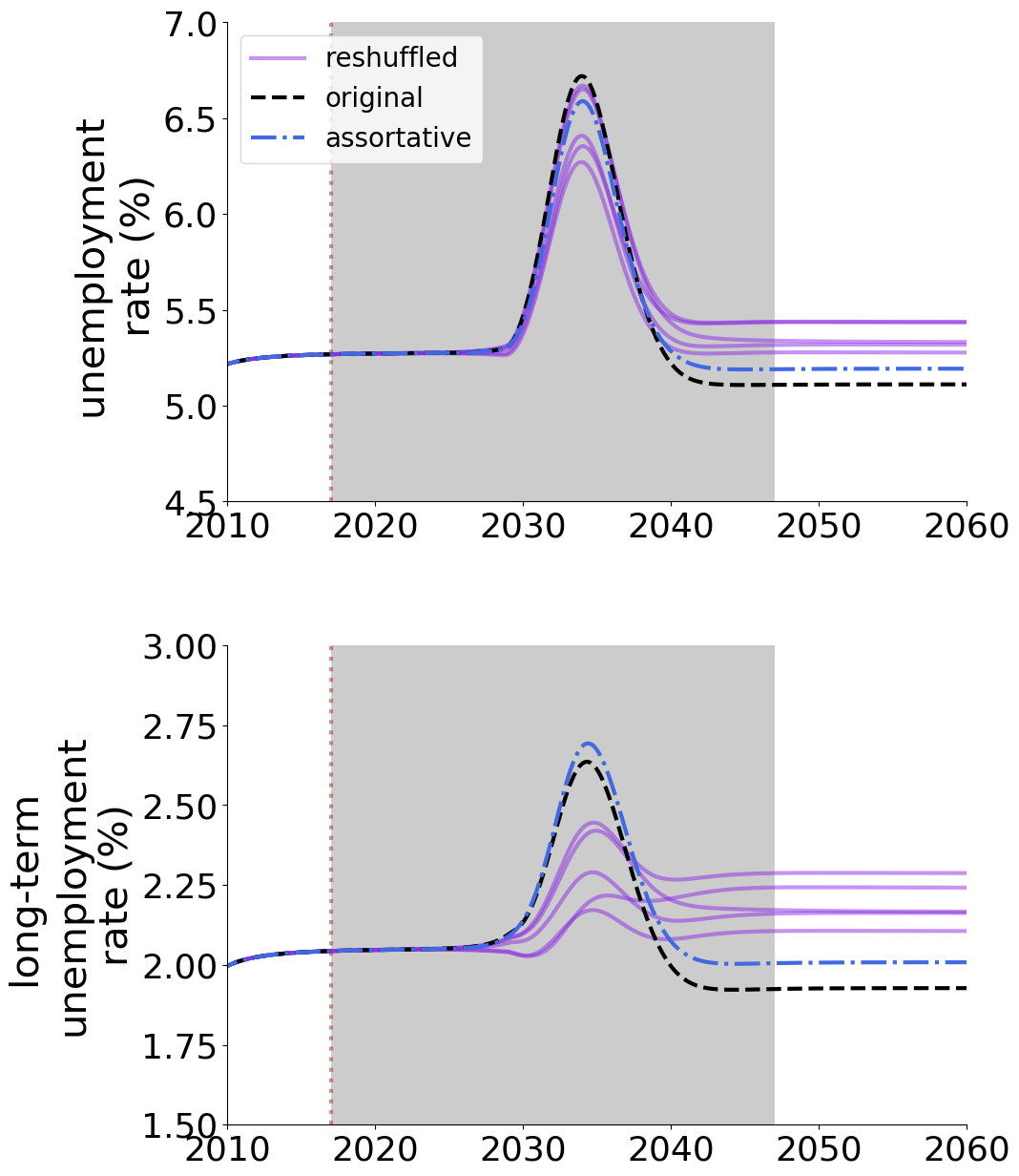}
\caption{\textbf{Randomized and assortative versions of the Probability of Computerization shock.} On the top the unemployment rate. On the bottom the long-term unemployment rate.}
\label{fig:rand_shock_FO}
\end{center}
\end{figure}

In Fig. \ref{fig:rand_shock_FO} we show our results for the original Probability of Computerization shock, an assortative reordering and five different randomized reshuffling. We observe that all random reshufflings have a higher unemployment rate at the steady-state, while the original and the assortative versions have a lower unemployment rate at the post-automation steady-state. The same is true for the long-term unemployment.

\section{Additional mathematical analysis}
In this section we discuss additional properties of the model and particular cases that may of interest to the reader.

\subsection{Dynamics of the realized demand}\label{sec:dyn_real_demand}
We showed that when the target demand is constant there exists a steady-state solution for $\bar{d}_{i,t}$. In this section we discuss the dynamics of $d_{i,t}$ and show that $d_{i,t}$ fluctuates around $d_i^*$. 

First, we note that $\omega_{i,t}$ and $\nu_{i,t}$ are binomial random variables of $e_{i,t}$ draws and success probability $\pi_{u,i,t}$ and $\pi_{u,i,t}$ respectively. Therefore, in the limit fo a large number of agents their distributions are the normals,
\begin{equation}
    \omega_{i,t} \sim \mathcal{N}(e_{i,t}\pi_{u,i,t}, \pi_{u,i,t}(1 - \pi_{u,i,t})e_{i,t})
    \label{eq:omega_approx}
\end{equation}
\begin{equation}
    \nu_{i,t} \sim \mathcal{N}(e_{i,t}\pi_{v,i,t}, \pi_{v,i,t}(1 - \pi_{v,i,t})e_{i,t}).
    \label{eq:nu_approx}
\end{equation}
Their difference is distributed as follows
\begin{equation}
    \nu_{i,t}  -  \omega_{i,t} =  e_{i,t}( \pi_{v,i,t} - \pi_{u,i,t}) + \phi_{i, t}
\end{equation}
where
\begin{equation}
  \phi_{i,t} \sim \mathcal{N}\left(0, \left( \pi_{u,i,t}(1 - \pi_{u,i,t}) + \pi_{v,i,t}(1 - \pi_{v,i,t}) \right)e_{i,t} \right).   
\end{equation}
Then, from Eqs. (6--8) 
we obtain that
\begin{equation}
  d_{i,t + 1}  = d_{i,t}  + (\delta_v - \delta_u)\bar{e}_{i,t}  + \gamma_u'(d^\dagger_{i,t} -  \bar{d}_{i,t}) + (\gamma_v' - \gamma_u') \max \big\{0, d_{i,t}^\dagger -  \bar{d}_{i,t} \big\} + \phi_{i,t}
  \label{eq:d_t_exact0}
\end{equation}
When $\delta_u > \delta v$, it follows from Eq. (\ref{eq:d_*}) that 
$$ d_i^\dagger = d_i^* +  \frac{\delta_u - \delta_v}{\gamma_v'} e_{i,t}. $$
We introduce this expression in Eq. \ref{eq:d_t_exact0} and obtain that
\begin{equation}
  d_{i,t + 1}  =\begin{cases}
(1 - \gamma_u')d_{i,t} - \gamma_u'd_i^* - (\delta_u - \delta_v)(1 - \frac{\gamma_u'}{\gamma_v'})e_{i,t} + \phi_{i, t+ 1}  & \text{if } \bar{d}_{i,t} \geq d^\dagger_{i,t}  \\
  (1 - \gamma_v')d_{i,t} - \gamma_v'd_i^*  + \phi_{i, t+ 1} & \text{if } \bar{d}_{i,t}< d^\dagger_{i,t}.
  \end{cases}
  \label{eq:d_t_exact}
\end{equation}
In the second case, $d_{i,t + 1}$ follows AR(1) process with coefficient $\gamma_v'$. On the first case, $d_{i,t + 1}$ follows an AR(1) process with and additional negative term that makes $d_{i,t + 1}$ get closer to the value of $d_i^*$. Therefore, $d_{i,t}$ fluctuates around $d_i^*$. In the particular case when $\delta_u = \delta_v = \delta$ and $\gamma_u = \gamma_v = \gamma$ we obtain the canonical AR(1) process equation
$$d_{i, t+1} = (1 - \gamma)d_{i,t} + \gamma d_i^* + \phi_{i, t+ 1}. $$

\subsection{Zero steady-state for the Agent-Based model}\label{sec:SI-zero-steady_state_abm}
Eqs (2--4) accept a trivial steady-state $e_i^* = u_i^* = v_{i}^* = d_{i}^\dagger = 0$. We neglect this steady-state since it is uninteresting for our analysis. However, when running the agent simulation there is a non-zero probability that the employment of an occupation is zero, i.e.,  $e_{i,t} = v_{i,t} = 0$. At this point, even if $d_{i,t}^\dagger  > 0$, no vacancies would open and therefore employment would be zero for the rest of the simulation. Therefore, we introduce the additional rule that if $e_{i,t} = v_{i,t} = 0$ but $d_{i,t}^\dagger  > 0$, then a vacancy is opened with probability $1$. 
When running the simulation with a large number of agents (which is the case of the labor market) the probability that $e_{i,t} = v_{i,t} = 0$ is negligible and this additional rule is not used.

\subsection{Steady-state for the complete network}\label{sec:SI-math_ucomplete}
We now show that our model has an analytically computable steady-state under the following assumptions: i) a complete network of $n$ nodes, i.e., $A_{ij} = \frac{1}{n} \quad \forall i,j$, ii) $\delta_u = \delta_v = \delta$ and that $\gamma_u = \gamma_v = \gamma$, and iii) the target labor demand is constant, equal to the labor supply, and distributed homogenously among all occupations i.e.,  $d_i^\dagger = \frac{L}{n} \quad \forall i$. As before, we denote the steady-state value of the variables with a star superindex (e.g. $x^*$).

In the main text we show that the steady-state depends on the target demand and the network structure. Since in this scenario all occupations have equal target demand and are positioned indistinguishably in the network, all occupations have the same steady state. Therefore, we lose the $i$ subindex in our notation. 
Since $\delta_u = \delta_v$,  Eq. (\ref{eq:u_*}) yields $ \sum_j \bar{f}_{ji}^* = \sum_j \bar{f}_{ij}^* \equiv F^*$. Using the full expression for the flow of workers in Eq.(\ref{eq:f_full}) we obtain

$$F = \sum_{j=1}^n \frac{1}{n}\frac{u^*v^{*2} (1 - e^{s^*/v^*}) }{s^* \sum_{k=1}^n\frac{1}{n}v^*} = \frac{u^*v^*(1-e^{-s^*/v^*})}{s^*}$$

similarly, from Eq.8
we have that
$$ s^* = \sum_{i=1}^n \frac{1}{n}\frac{u^*v^*}{\sum_{k=1}^n\frac{1}{n} v^*} = u^*.$$
We then substitute $s^*$ in $F$ and obtain,
\begin{equation}
 F = v^*(1-e^{-u^*/v^*})
 \label{eq:F}
\end{equation}

It follows from assumption ii) and Eq.(\ref{eq:d_*}) that at the steady state  $\bar{d}^* = d^\dagger$. With this in mind, using Eq.(\ref{eq:F}) and Eqs.13--15
, we obtain the following dynamic equations for the total number of employed and unemployed workers and job vacancies, 
\begin{equation}
    e^* = e^* - \delta e^*  + v^*(1-e^{-u^*/v^*})
    \label{eq:complete_e*}
\end{equation}
\begin{equation}
    u^* = u^* + \delta e^*  - v^*(1-e^{-u^*/v^*})\nonumber
\end{equation}
\begin{equation}
    v^* = v^* + \delta e^*  - v^*(1-e^{-u^*/v^*}).\nonumber
\end{equation}
We know that the number of unemployed and employed workers equals the labor, i.e., $U^* + E^* = L$. Since all occupations have the same steady state, then $u^* + e^* = \frac{L}{n}$. It follows from this observation, the fact that $d^* = \frac{L}{n}$, and Eq. \ref{eq:d_*} that
\begin{equation}
    u^* = \frac{L}{n} - e^* = v^*.
    \label{eq:steady_uev}
\end{equation}
Finally, we substitute Eq.(\ref{eq:steady_uev}) into Eq.(\ref{eq:complete_e*}) and obtain

\section{Additional tables and datasets}

\begin{longtable}{p{3.0cm}|p{10cm}}
\textbf{Main variables}             & \textbf{Description}                                    \\
\hline
$e_{i, t}$            & Number of employed workers at time $t$ on occupation $i$.  \\
$u_{i, t}$            & Number of unemployed workers at time $t$ which were last employed in occupation $i$ \\
$v_{i, t}$            & Number of job vacancies at time $t$ of occupation $i$ \\
& \\
\textbf{Other variables}             &                                   \\
\hline
$d_{i, t} $   & Realized labor demand at time $t$ of occupation $i$. ($d_{i, t} = e_{i, t} + v_{i, t}$) \\
$q_{ij, t}$ & The probability that an unemployed worker from occupation $i$ applies to a job vacancy of occupation $j$ at time $t$.\\
$p_{j, t} $ & The probability that a job application sent to a vacancy of occupation $j$ is successful.\\
$s_{ij, t} $ & The number of job applicants occupation $j$ receives from workers of occupation $i$ at time $t$. \\
$s_{j, t} $ & The number of job applicants occupation $j$ receives at time $t$. \\
$f_{ij, t}$ & Flow of unemployed workers of $i$ to be employed at occupation $j$\\
$u_{i, t}^{(k)} $ & The number of unemployed workers of occupation $i$ that have spent exactly $k$ time steps unemployed at time $t$.  \\
$\pi_{u,i,t} $ & Occupation-specific probability that a worker of occupation $i$. Though not noted explicitly, this probability is time dependent. \\
$\pi_{v,i,t} $ & Occupation-specific probability that a vacancy of occupation $i$ opens, per worker employed in occupation $i$. Though not noted explicitly, this probability is time dependent. \\
$\omega_{i, t}$        & Number of workers from occupation $i$ that separated from their jobs at time $t$. Drawn from Binomial distribution $Bin(e_{i,t}, \pi_{u,i,t})$. \\ 
$\nu_{i, t}$             & Number of vacancies of occupation $i$ opened at time $t$. Drawn from Binomial distribution $Bin(e_{i,t}, \pi_{v,i,t})$  \\ 
& \\
\textbf{Parameters}  &  \\
\hline
$\delta_u$              & Rate at which employed workers are separated due to the spontaneous process. \\ 
$\delta_v$              & Rate at which employed vacancies are opened due to spontaneous process. \\ 
$\gamma$              & Rate at employed workers and vacancies are separated or opened due to the market adjusting towards the target demand.                \\
$\tau$ & Number of time steps after which an unemployed workers if considered long-term unemployed\\
$r$ & weights of the self-loops of the occupational mobility network. \\
$A$ & Adjacency matrix of the occupational mobility network \\
$d^\dagger_{i}$             & Post-automation target labor demand. Number of workers needed at occupation $i$ after the automation shock is complete. \\
$d^\dagger_{i,t}$& Target labor demand of occupation $i$ at time $t$\\
$\Delta t$& Duration of a time step in units of weeks.\\
\caption{Variables and parameters}
\label{tab:variables_and_parameters}
\end{longtable}








\subsection*{Table S2: Results for all occupations for the Frey and Osborne shock}

\begin{tiny}
\begin{longtable}{p{4.5cm}|p{1.5cm}|p{1.5cm}|p{1.5cm}|p{1.5cm}|p{1.5cm}}
Occupation & Probability of computerization & Percentage change in the average unemployment rate & Percentage change in the average long-term unemployment rate & Unemployment rate before shock & Target demand before shock \\
\hline
  machine tool cutting setters, operators, and t... &                      0.90 &                       62 &                  297 &                       2.8 &           154172 \\
                             postal service clerks &                      0.95 &                       65 &                  238 &                       2.8 &           114948 \\
 extruding, forming, pressing, and compacting m... &                      0.93 &                       73 &                  237 &                       3.2 &            33072 \\
 woodworking machine setters, operators, and te... &                      0.97 &                       81 &                  203 &                       3.5 &            18782 \\
                    machine feeders and offbearers &                      0.93 &                       72 &                  195 &                       3.6 &            27175 \\
    adhesive bonding machine operators and tenders &                      0.95 &                       87 &                  192 &                       3.7 &             8809 \\
             automotive body and related repairers &                      0.91 &                       58 &                  186 &                       3.2 &           136839 \\
\end{longtable}
\end{tiny}

\newpage

\subsection*{Table S3: Results for all occupations for the Brynjolfsson et al. shock}
\begin{tiny}
\begin{longtable}{p{4.5cm}|p{1.5cm}|p{1.5cm}|p{1.5cm}|p{1.5cm}|p{1.5cm}}
Occupation & Probability of computerization & Percentage change in the average unemployment rate & Percentage change in the average long-term unemployment rate & Unemployment rate before shock & Target demand before shock \\
\hline
 drafters &  3.90 &                        6 &                   28 &                       2.7 &           164566 \\
         baggage porters, bellhops, and concierges &  3.80 &                        6 &                   19 &                       3.5 &            91661 \\
           appraisers and assessors of real estate &  3.70 &                        4 &                   16 &                       3.0 &            93101 \\
                             postal service clerks &  3.66 &                        3 &                   16 &                       2.8 &           114948 \\
                      postal service mail carriers &  3.65 &                        3 &                   14 &                       3.1 &           331040 \\
    morticians, undertakers, and funeral directors &  3.89 &                        5 &                   12 &                       3.8 &            38463 \\
                                     travel agents &  3.66 &                        4 &                   12 &                       3.9 &            69566 \\
\end{longtable}
\end{tiny}


\end{document}